\documentclass[12pt,letterpaper]{article}

\usepackage{amsmath,amssymb,array,calc,rotating,epsfig,psfrag}
\usepackage[nosort]{cite}
\usepackage{graphicx}
\usepackage{hyperref}
\usepackage{mathtools,slashed}
\usepackage[utf8]{inputenc} 
\usepackage[dvipsnames]{xcolor}
\usepackage{tensind}
\usepackage{tensor}
\usepackage{cleveref}
\tensordelimiter{?}
\usepackage{float} 
\usepackage{subcaption} 
\usepackage{multirow} 
\usepackage{url}

\numberwithin{equation}{section}

\def\a{\alpha}
\def\b{\beta}

\def\m{\mu}
\def\g{\gamma}
\def\l{\lambda}
\def\n{\nu}

\def\t{\tau}

\def\a{\alpha}
\def\b{\beta}
\def\d{\delta}
\def\g{\gamma}
\def\G{\Gamma}

\def\rd{{\rm d}}

\def\cD{{\mathcal{D}}}
\def\cP{{\mathcal{P}}}
\def\cK{{\mathcal{K}}}

\def\tG{ {\tilde \G}}

\def\TCN{{$\text{TC}\mathcal{N}$}}

\def\be{\begin{equation}} 
\def\ee{\end{equation}}

\newcommand*{\figtitle}[1]{{\centering\footnotesize{\text{#1}}\par\medskip}}

\begin{document}

\begin{titlepage}
\begin{center}
\baselineskip .9 cm
{\Large \bf Inflation from Dynamical Projective Connections}\\
\vskip 0.4 cm
\baselineskip .4 cm
{\bf \footnotesize  Muhammad Abdullah$^{a}$, Calvin Bavor$^{d,}$\footnote{cbavor62@gmail.com},  
Biruk Chafamo$^{a}$, Xiaole Jiang$^{a}$, Muhammad Hamza Kalim$^{a}$,
Kory Stiffler$^{b,c,}$\footnote{kory-stiffler@uiowa.edu}, and Catherine A. Whiting$^{a,d,}$\footnote{cwhiting@coloradomesa.edu}}

\vspace{.4cm}

{\it ${}^a$ Department of Physics and Astronomy}\\
{\it Bates College, Lewiston, ME 04240}

\vspace{.2cm}


\vspace{.2cm}

{\it ${}^b$ Department of Physics and Astronomy}\\
{\it   The University of Iowa,   Iowa City, IA 52242, USA}\\

\vspace{.2cm}

{\it ${}^c$ Brown Theoretical Physics Center and Department of Physics}\\
{\it Brown University, Providence, RI 02912-1843, USA }

\vspace{.2cm}

{\it ${}^d$ Department of Physical and Environmental Sciences}\\
{\it Colorado Mesa University, 1100 North Avenue, Grand Junction, CO 81501, USA }
\end{center}
\vspace{.2cm}

\begin{abstract}
We show how the recently developed string-inspired, projectively-invariant gravitational model Thomas-Whitehead gravity (TW gravity) naturally gives rise to a  field acting as the inflaton. In the formulation of TW gravity, a field $\mathcal{D}_{ab}$ is introduced into the projective connection components and is related to a rank-two tensor field $\mathcal{P}_{ab}$. Through the dynamical action of TW gravity, in terms of projective curvature, the tensor field $\mathcal{P}_{ab}$ acquires dynamics. By decomposing $\mathcal{P}_{ab}$ into its trace and traceless degrees of freedom, and choosing the connection to be Levi-Civita, we demonstrate that TW gravity contains a non-minimally coupled scalar field with a specific potential. Considering only the trace degrees of freedom, we demonstrate that the scalar field acts as an inflaton in the slow roll approximation. We find a range of values for the parameters introduced by TW gravity that fit the experimental constraints of the most recent cosmological data. 

\end{abstract}

\end{titlepage}

\section{Introduction}
  Since the initial formulation of cosmological inflation in the late 1970's to early 1980's \cite{Starobinsky1979,Guth1981,Linde1982,Albrecht1982}, dynamical scalar fields representing the so-called inflaton field have appeared in many unique forms. Despite the overwhelming evidence that the inclusion of scalar fields both alleviates long-standing cosmological problems and predicts the observed nearly scale-invariant spectrum of CMB perturbations, there are few proposals for a fundamental physical origin of the inflaton. Indeed, as explained by Kolb and Turner
      \begin{quote}
        \emph{That paradigm [inflation], however, is still without a standard model for its implementation. Of course, that shortcoming should be viewed in light of the fact that our understanding of physics at energy scales well beyond that of the standard model of particle physics is still quite incomplete.}
      \end{quote}
      \hspace*{60 pt}-E. Kolb and M. Turner~\cite{kolbt}
      \vspace*{6 pt}
      
Due to the expectation that physics beyond the standard model should have something to say about inflation, substantial effort has been focused on teasing out the emergence of inflation through string theory motivated models \cite{baumann_mcallister_2015}. Attempts have also been made to realize the inflaton as the Standard Model Higgs field \cite{Bezrukov2008,Rubio2019}. Here, the Higgs is only a potentially viable inflaton if it couples non-minimally to the gravitational sector with a coupling of the form $\xi \phi^2 R$. More general models of non-minimally coupled inflation consider various other forms for the potential and coupling \cite{Fakir1990,Watanabe2007,Park:2008hz,Pallis2010,kodama2021relaxing}.

Thomas-Whitehead gravity (TW gravity) \cite{Brensinger:2017gtb,Brensinger:2020mnx,Brensinger:2020gcv} is a string-inspired model of gravity that emerges from the projective geometry of Thomas and Whitehead \cite{Thomas:1925a,Thomas:1925b,Whitehead:1931a}. In this paper, we demonstrate this model could be the aformentioned raison d'etre for inflation, the foundational principle being projective symmetry. The importance of projective symmetry arises upon extending a coadjoint element of the Virasoro algebra to higher dimensions~\cite{Rai:1989js,Delius:1990pt,Lano:1994gx,Branson:1996pe,Branson:1998bc,Rodgers:2003an,Rodgers:2006ep,Brensinger:2017gtb,Brensinger:2020gcv}. That the ultimate foundational piece of TW gravity is the Virasoro algebra is why TW gravity is said to be \emph{string-inspired}. We find solutions to TW gravity that describe an early universe inflationary epoch fitting current cosmological data~\cite{PCX2019}. These solutions are parameterized by a set of three fundamental constants in TW gravity. One choice of these parameters would correspond to a certain combination of scalar field (non-minimally coupled) inflationary models investigated in~\cite{kodama2021relaxing}.

This paper is structured as follows. In section~\ref{s:ReviewB} we review TW gravity. We demonstrate how projective symmetry is utilized in the construction of TW gravity and  summarize the connection to the deeper underlying Virasoro algebra. In section~\ref{s:NMCInflation} we demonstrate how a non-minimally coupled (NMC) model for inflation with a specific potential naturally emerges from TW gravity.  This NMC model has three free parameters inherited from the full TW gravity model. 
In the slow roll approximation we constrain these three parameters within a range that matches the current observational data for the spectral index $n_s$, tensor-to-scalar ratio $r$, and scalar amplitude $A_s$ for efoldings of $N=50$, $60$, and $70$. In section~\ref{s:conclusion} we make concluding remarks. The appendices show our conventions, reviews of background material, and supporting calculations.

\section{Review of Thomas-Whitehead Gravity}\label{s:ReviewB}
In this section we review TW gravity by defining projective connections, building from this projective curvature, and from this constructing projective invariants. We then show how the projectively invariant action of TW gravity is composed of these projective invariants.

\subsection{Projectively Equivalent Paths and Projective Connections}\label{s:PathsAndConnections}
Here we briefly review the necessary components of TW gravity, as generalized recently in~\cite{Brensinger:2020gcv} from the constant volume form of \cite{Brensinger:2017gtb,Brensinger:2020mnx}. TW gravity is a theory of dynamical projective connections. For a detailed review of the theory of projective connections, we refer the reader to \cite{Crampin}. A connection $\nabla_a$ with coefficients $\G^a{}_{bc}$ describes a geodesic path with coordinates $x^a$ and parameterization $\tau$. These satisfy the geodesic equation
\begin{align}\label{e:GeodesicEq}
    \frac{ dx^b}{d\tau} \nabla_b \frac{d x^a}{d\tau} \equiv \frac{d^2 x^a}{d\tau^2} + \G^a{}_{bc} \frac{d x^b}{d\tau} \frac{d x^c}{d\tau} = 0\;.
\end{align}
The same geodesic path with coordinates $x^a$ can be described by a different connection $\hat{\nabla}_a$ with coefficients $\hat{\Gamma}^a{}_{bc}$ and reparameterization $\sigma = \sigma(\tau)$. These also satisfy the geodesic equation
\begin{align}\label{e:GeodesicEqHat}
     \frac{ dx^b}{d\sigma} \hat{\nabla}_b \frac{d x^a}{d\sigma} \equiv  \frac{d^2 x^a}{d\sigma^2} + \hat{\G}^a{}_{bc} \frac{d x^b}{d\sigma} \frac{d x^c}{d\sigma} = 0\;,
\end{align}
so long as the connection coefficients and parameterizations are related as follows
\begin{align}
\label{e:proj_trans}
\Gamma\indices{^a _{bc}}=&\hat{\Gamma}\indices{^a _{bc}}+\delta \indices{^a _b}v_c +\delta\indices{^a _c}v_b\;, \\
\label{e:Reparameterization}
\frac{d^2 \sigma}{d\tau^2} = & - 2 \left( \frac{d\sigma}{d\tau}\right)^2 \frac{d x^a}{\d\sigma}v_a \;.
\end{align} 
Equation~\eqref{e:proj_trans} is known as a projective transformation, with $v_a$ an arbitrary one-form.. That Eqs.~\eqref{e:GeodesicEq} and~\eqref{e:GeodesicEqHat} take the same form demonstrates the projective equivalence of the connections $\nabla_a$ and $\hat{\nabla}_a$.

The theory of projective connections seeks to make this equivalence of paths under reparameterizations into a manifest symmetry. Given the base spacetime $\mathcal{M}$, where the spacetime connection $\nabla_a$ is defined, projective connections are defined in the space of one higher dimension.  We refer to this $\rd+1$-dimensional space as the Thomas-Cone $\mathcal{N}$.  Like the usual spacetime connection  $\nabla_a$, the projective connection can be used to form curvature invariants and, through these curvature invariants, one can write dynamical actions which are invariant under both projective transformations and general coordinate transformations.

We now seek to make explicit the construction of the projective connection. In the following, Latin indices $a,b,c,\ldots=0,1,\ldots,\rd-1$ are reserved for the $\rd$-dimensional spacetime $\mathcal{M}$ coordinates $x^a$ and Greek indices $\alpha, \beta,\gamma,\ldots=0,1,\dots \rd$, excluding $\lambda$, are reserved for the $\rd+1$-dimensional Thomas-Cone $\mathcal{N}$ coordinates $x^{\alpha}$. The index $\lambda$ refers to the extra coordinate $x^{\rd}=x^\lambda\equiv \lambda$ of the Thomas-Cone $\mathcal{N}$. The coefficients of the projective  connection $ {\tilde \nabla}_\a $ are defined by requiring that
\be {\tilde \nabla}_\a \Upsilon^\b = \d_\a{}^\b \; ,
\ee
where $\Upsilon$ is the fundamental vector field of the Thomas-Cone $\mathcal{N}$ generating projective transformations. Explicitly, the projective connection coefficients are defined by
\begin{equation} 
{{\tilde\G}}^{\a}_{\,\,\b \g}= \begin{cases}
    {\tilde \G}^{\lambda}_{\,\,\,\lambda a}={\tilde \G}^{\lambda}_{\,\,\, a \lambda} = 0
    \\ {\tilde \G}^{\a}_{\,\,\,\,\lambda \lambda} = 0 \label{e:Gammatilde}\\ {\tilde \G}^{a}_{\,\,\,\,\lambda b}={\tilde \G}^{a}_{\,\,\,\,b \lambda} = \a_\lambda\,\d^a_{\ b}\\
{\tilde \G}^{a}_{\,\,\,\,b c} ={ \Pi}^{a}_{\,\,\,\,b c}\\
{\tilde \G}^{\lambda}_{\,\,\,\, a b} =  \Upsilon^\lambda \cD_{ a b}
 \end{cases}  \; ,
\end{equation} 
where 
\begin{align}
\label{e:Pi}
{\Pi}^{a}_{\,\,\,\,b c} =& { \G}^{a}_{\,\,\,\,b c} + \delta^a{}_{(c}~ \a_{b)}~~~,~~~\a_a = -\tfrac{1}{\rd +1} \Gamma^m{}_{am}\\
        \Upsilon^\rho =& (0,0,\dots,0,\lambda)~~~,~~~\a_\rho = \left( \a_a, \lambda^{-1} \right). \label{UpsilonOmega}
\end{align}
In Eq.~\eqref{e:Gammatilde} to Eq.~\eqref{UpsilonOmega}, $\Gamma\indices{^a _{bc}}$ are the connection coefficients  on the the $\rd$-dimensional spacetime $\mathcal M$. The object $\Pi\indices{^a _{bc}}$ appearing with all indices on $\mathcal{M}$ is known as the fundamental projective invariant. This object is invariant under the projective transformation Eq.~\eqref{e:proj_trans}.  Thus, $\Pi\indices{^a _{bc}}$ determines the equivalence classes of projectively related connections. Note that at this point the spacetime connection $\Gamma \indices{^a _{bc}}$ is not necessarily constrained to be the Levi-Civita connection, though we will later make this assumption. Instead, $\Gamma \indices{^a _{bc}}$ is a representative of the projective equivalence class of connections $[\Gamma \indices{^a _{bc}}]$, which are equivalent under projective transformations and reparameterization as in Eqs.~\eqref{e:proj_trans} and~\eqref{e:Reparameterization}. The one-form  $\alpha$ is related to $\Upsilon$ by the conditions that  $\alpha _\rho \Upsilon^\rho=1  $ and $\mathfrak{L}_\Upsilon \alpha_\rho =0$, where $\mathfrak{L}_\Upsilon$ denotes the Lie derivative with respect to $\Upsilon$. 

The symmetric field $\mathcal{D}_{ab}$, known as the diffeomorphism field, is not a tensor due to its appearance as part of the connection coefficient $\tilde{\Gamma}\indices{^\lambda _{ab}}$. It transforms as follows
\begin{align}\label{e:Dtrans}
    \mathcal{D}'_{ab} =& \frac{\partial x^m}{\partial x'^a}\frac{\partial x^n}{\partial x'^b} \left[\mathcal{D}_{mn} - \partial_m j_n - j_m j_n + j_c \Pi^c{}_{mn}\right] \;, \\
    j_m \equiv & \partial_m  \log | \tfrac{\partial x^b}{\partial x'^c} |^{\tfrac{1}{\rd+1}} \;,
\end{align}
with $| \tfrac{\partial x^b}{\partial x'^c} | $ the Jacobian of the transformation. For an infinitesimal coordinate transformation $x' = x - \xi(x)$ in $\rd =1$-dimension, the transformation Eq.~\eqref{e:Dtrans} becomes
\begin{align}\label{e:Dtrans1D}
    \mathcal{D}'(x) = \mathcal{D}(x) + 2 \frac{d \xi(x)}{dx}\mathcal{D}(x) + \xi(x) \frac{d \mathcal{D}(x)}{d x} - \frac{1}{2} \frac{d^3\xi(x)}{dx^3} \;,
\end{align}
to first order in $\xi$. The details of this dimensional reduction are shown in appendix~\ref{a:DtransDetails}. Equation~\eqref{e:Dtrans1D} is the same transformation law as a coadjoint element $\mathcal{D}$ of the Virasoro algebra, up to rescalings of $\mathcal{D}$. The diffeomorphism field $\mathcal{D}_{ab}$ is usually seen as the dynamical extension of a coadjoint element of the Virasoro algebra to higher dimensions, rather than the dimensional reduction that was described here.  From this vantage point, TW gravity as the resulting dynamical model for $\mathcal{D}_{ab}$ is said to be \emph{string-inspired} from the core foundational Virasoro algebra in one-dimension. More details can be found in the seminal works~\cite{Rai:1989js,Delius:1990pt,Lano:1994gx,Branson:1996pe,Branson:1998bc,Rodgers:2003an,Rodgers:2006ep,Brensinger:2017gtb,Brensinger:2020gcv}.

Once in arbitrary $\rd$-dimensions, it is advantageous to form a tensor by adding various objects to $\mathcal{D}_{bc}$ that remove the non-tensorial pieces from its transformation law. This is accomplished through the definition of the tensor field $\mathcal{P}_{bc}$
\begin{align}
\label{e:PtoD}
                \mathcal{P}_{bc} = \mathcal{D}_{bc} - \partial_b \a_c + \Gamma^e_{\ bc}\a_e + \a_b \a_c \;.
\end{align} 
Note that $\mathcal{P}_{bc}$ is symmetric only if the curl of $\alpha_c$ vanishes, one possible solution being that the spacetime connection is Levi-Civita with respect to a metric $g_{ab}$ on $\mathcal{M}$ such that $\Gamma\indices{^m _{am}}=\partial_a \ln \sqrt{|g|}$.
This field $\mathcal{P}_{bc}$, related to the diffeomorphism field $\mathcal{D}_{bc}$ via Eq.~\eqref{e:PtoD}, transforms as a tensor on the spacetime $\mathcal{M}$ and will be shown to act as a source of cosmological inflation under certain assumptions. 

Under a general coordinate transformation on $\mathcal{M}$, the spacetime connection $\G^a{}_{bc}$ transforms as an affine connection
\begin{align}\label{e:GTrans}
        {\G}'^{a}{}_{mn} =& \frac{\partial x'^a}{\partial x^b}\frac{\partial x^p}{\partial x'^m}\frac{\partial x^q}{\partial x'^n}{\G}^{b}{}_{pq} + \frac{\partial^2 x^b}{\partial x'^m \partial'^n}\frac{\partial x'^a}{\partial x^b}\;.
\end{align}
The projective connection~$\tG^\a{}_{\m\n}$ transforms as an affine connection 
\begin{align}\label{e:tGTrans}
        \tG'^{\alpha}{}_{\mu\nu} =& \frac{\partial x'^\a}{\partial x^\rho}\frac{\partial x^\sigma}{\partial x'^\m}\frac{\partial x^\beta}{\partial x'^\n}\tG^{\rho}{}_{\sigma\beta} + \frac{\partial^2 x^\beta}{\partial x'^\mu \partial'^\nu}\frac{\partial x'^\a}{\partial x^\b}
\end{align}
under what we refer to as a Thomas-Cone transformation on $\mathcal{N}$ (\TCN-transformation)~\cite{Thomas:1925a, Thomas:1925b, Roberts, Brensinger:2017gtb, Brensinger:2020mnx, Brensinger:2020gcv}
\begin{align}
        \label{e:ProjTrans}
        x'^\a =& ( x'^0(x^m), x'^1(x^m), \dots , x'^{\rd -1}(x^m) , \lambda' = \lambda J^{1/(\rd+1)} )~~~,~~~ J = |\partial x^m / \partial x'^n| \; .
\end{align}
The TC$\mathcal{N}$-transformation is seen to be a general coordinate transformation on $\mathcal{M}$ with an additional Jacobian scaling of the $\lambda$-direction. We refer to objects transforming as a tensor with respect to TC$\mathcal{N}$-transformations as TC$\mathcal{N}$-tensors. Now we have the technology to write the manifestly TC$\mathcal{N}$-covariant and manifestly projectively invariant geodesic equation
\begin{align}\label{e:GeodesicEqTCN}
     \frac{ dx^\m}{d\tau} \tilde{\nabla}_\m \frac{d x^\a}{d\t} \equiv  \frac{d^2 x^\a}{d\t^2} + \tilde{\G}^\a{}_{\m\n} \frac{d x^\m}{d\t} \frac{d x^\n}{d\t} = 0\;.
\end{align}
This equation is manifestly TC$\mathcal{N}$-covariant in that $dx^\mu/d\tau$ and $\tilde{\nabla}_\mu$ are both TC$\mathcal{N}$-tensors. At the same time, it is also manifestly projectively invariant in that $\tilde{\Gamma}^\a{}_{\m\n}$ is invariant with respect to projective transformations as in Eq.~\eqref{e:proj_trans}.

\subsection{Projective Curvature}
Since the projective connection coefficients $\tilde{\Gamma}\indices{^\alpha _{\mu \nu}}$ transforms as an affine connection under a TC$\mathcal{N}$ transformation, we can  straightforwardly compute its curvature invariants.  Explicitly, on a vector field \(\kappa^\a\) and co-vector \(\kappa_\a\) in \(\mathcal N,\) we define the projective curvature tensor ${\cK}^{\g}_{\,\,\,\rho\a \b} $ through the usual relations
\be 
[{\tilde \nabla}_\a,{\tilde \nabla}_\b] \kappa^\g = \,{\cK}^{\g}_{\,\,\,\rho\a \b  } \kappa^\rho~~~,~~~[{\tilde \nabla}_\a,{\tilde \nabla}_\b] \kappa_\g =- \,{\cK}^{\rho}_{\,\,\,\g\a \b  } \kappa_\rho \; .
\ee
In terms of the connection coefficients, we can write the projective curvature as
\begin{align}\label{e:PCurv}
        \cK^\mu{}_{\nu\alpha\b} \equiv \tilde{\G}^\mu{}_{\nu[\b,\a]} + \tilde{\G}^\rho{}_{\n[\b}\tilde{\G}^\mu{}_{\a]\rho} \; .
\end{align}

The extended metric $G_{\a \b}$ on the $(\rd+1)$-dimensional manifold $\cal{N}$ is written succinctly as
\begin{align}\label{e:BigGSuccinct}
        G_{\a\b} &= \delta^a_{\,\,\alpha} \delta^b_{\,\,\beta} \,g_{ab} - \lambda_0^2 g_\alpha g_\b\\
        G^{\a\b} &= g^{ab} (\delta^\alpha_{\,\,a} - g_a \Upsilon^\a)(\delta^\b_{\,\,b} - g_b \Upsilon^\b) - \lambda_0^{-2} \Upsilon^\a\Upsilon^\b,
\end{align} 
where $g_\alpha = (g_a,1/\lambda)$ and  $g_a\equiv -\frac{1}{\rd+1}\partial_a \ln\sqrt{|g|}$. This extension of a $\rd$-dimensional metric was detailed in \cite{Brensinger:2020gcv}, where the previous restriction of constant-volume coordinates was lifted.
The general volume coordinate metric $G_{\a\b}$ can be written as the sum of the constant volume metric $G_{\a\b}^{(0)}$ and finite correction $\Delta G_{\a\b}$:
\begin{align}
        G_{\a\b} = G_{\a\b}^{(0)} + \Delta G_{\a\b}~~~&,~~~~ G^{\a\b} = G^{\a\b}_{(0)} + \Delta G^{\a\b} \label{e:GmetricGen} \\
    G_{\a \b}^{(0)} = \left( 
                        \begin{array}{cc}
                                g_{ab} & 0 \\
                                0 & -\ell^{-2}
                        \end{array}
        \right)  ~~~&,~~~
        G^{\a \b}_{(0)} = \left( 
                        \begin{array}{cc}
                                g^{ab} & 0 \\
                                0 & -\ell^{2}
                        \end{array}
        \right) \label{e:GmetricAndInverse0} \\
        \Delta G_{\a \b} = \left( 
                        \begin{array}{cc}
                                -\lambda_0^2 g_a g_b & -\lambda_0\ell^{-1} g_a \\
                               -\lambda_0\ell^{-1}g_b & 0
                        \end{array}
        \right)  &~,~
        \Delta G^{\a \b} = \left( 
                        \begin{array}{cc}
                                0 & -\lambda g^{am}g_m \\
                                -\lambda g^{bm}g_m  & \lambda^2 g^{mn}g_m g_n
                        \end{array}
        \right) \label{e:deltaGmetricAndInverse} 
\end{align}
where $\ell \equiv \lambda/\lambda_0$.

 Note again that if the spacetime connection $\Gamma\indices{^a _{bc}}$ is the Levi-Civita connection then $g_a=\alpha_a$, since $\Gamma\indices{^a_{ab}}=\partial_b \ln\sqrt{|g|}$. The determinate of the metric $G_{\a\b}$ is the same as for $G^{(0)}_{\a\b}$, ~\cite{Brensinger:2017gtb,Brensinger:2020mnx}
\begin{align}
        G \equiv &\det (G_{\a\b}) = \det (G^{(0)}_{\a\b}) = -\ell^{-2} g~~~,~~~g \equiv \det(g_{ab})\;.
\end{align}

The  only non-vanishing components of $\cK^{\alpha}_{~\beta\mu\nu}$ are
\begin{align}
        \label{e:KRiemann}
        \cK^a{}_{bcd} =& R^a{}_{bcd} + \delta_{[c}{}^a \mathcal{P}_{d]b}-\delta \indices{^a _b}\mathcal{P}_{[cd]}~~~,\\ 
        \label{e:Kbreve}
      \breve{\mathcal{K}}_{nab}\equiv \alpha_\lambda \mathcal{\cK}^\lambda{}_{nab} =& \mathcal{P}_{n[b;a]} + \alpha_{[b}\mathcal{P}_{a]n}+\alpha_n\mathcal{P}_{[ab]} - R^{m}{}_{nab}\alpha_m~~~, 
       \\
        \label{e:KRicci}
        \cK_{ab} =& \cK^{\mu}{}_{a\mu b} =  R_{ab} +\rd \mathcal{P}_{ba}-\mathcal{P}_{ab} \\
        \label{e:K}
        \cK\equiv  & G^{\alpha\beta}\cK_{\alpha\beta} = R + (\rd - 1)\mathcal{P} \\
        \label{e:RD}
                R_{ab} \equiv R^m{}_{amb}~~~&,~~~R \equiv g^{ab}R_{ab}~~~,~~~ \mathcal{P} \equiv  g^{ab}\mathcal{P}_{ab}~~~.
\end{align}
\begin{equation}
R^{a}_{\ bcd}=\partial_c\G^a_{\ db}-\partial_d\G^a_{\ cb}+\G^a_{\ ce}\G^e_{\ db}-\G^a_{\ de}\G^e_{\ cb}
\end{equation}
The projective curvature tensor satisfies the following
\begin{align}
        \label{e:KRiemannSymmetries}
        \cK_{\alpha\beta\mu\nu} =& - \cK_{\alpha\beta\nu\mu}~~~ , ~~~  \cK_{\alpha[\beta\mu\nu]} = 0~~~.
\end{align}
An important object to consider will be the projective Cotton-York tensor
\begin{align}
	K_{\nu\a\b} \equiv g_{\mu} \cK^\mu{}_{\n\a\b}
\end{align}
which is a TC$\mathcal{N}$-tensor. Its only non-vanishing components are
\begin{align}\label{e:CYT}
	K_{nab} =&  \nabla_{[a}\cP_{b]n} -\Delta_n \cP_{[ab]} + \Delta_{[a}\cP_{b]n} + \Delta_{m}R^{m}{}_{nab} \\
	\Delta_a =& g_a - \alpha_a
\end{align}
Notice on the Levi-Civita shell, $\Delta_a = 0$ and the Cotton-York tensor becomes simply
\begin{align}
    K_{nab} = \nabla_{[a}\cP_{b]n}~~~\text{with}~~~\Delta_a = 0
\end{align}
and can be thought of as a gravitational analog of an electromagnetic field strength. The existence of such an analogy is not surprising. The more general Yang-Mills theory can be birthed from the Kac-Moody algebra in an analogous fashion to how TW gravity is birthed from the Virasoro algebra as summarized in section~\ref{s:PathsAndConnections}. More details can be found in the seminal works~\cite{Rai:1989js,Delius:1990pt,Lano:1994gx,Branson:1996pe,Branson:1998bc,Rodgers:2003an,Rodgers:2006ep,Brensinger:2017gtb,Brensinger:2020gcv}.

\subsection{Projectively Invariant Action}\label{s:ProjectiveAction}
We now detail the dynamical action describing TW gravity. This action is built from the projective curvature invariants described in the previous section and appears as
\begin{align}\label{e:STWdplus1}
           S_{TW} =& -\tfrac{1}{2\tilde{\kappa_0}} \int d\ell\; d^{\rd} x\sqrt{|G|} (\cK + 2 \Lambda_0)   \cr
        &- \tilde{J_0} c\int d\ell\; d^{\rd} x\sqrt{|G|} \left[\cK^2 - 4 \cK_{\alpha\beta}\cK^{\alpha\beta} + \cK \indices{^\alpha _{\beta \mu \nu}}\cK \indices{_\alpha ^{\beta \mu \nu}}   \right]\;.
\end{align}
This first line of Eq.~\eqref{e:STWdplus1} includes a projective Ricci scalar and cosmological constant, mimicking the usual Einstein-Hilbert action. The second line is the projective Gauss-Bonnet action allowing for $\mathcal{P}_{ab}$ to acquire dynamics. Specifically, dynamics is given to $\mathcal{P}_{ab}$ through the $\breve{\mathcal{K}}_{nab}$ components of $\mathcal{K}^\a{}_{\b\m\n}$ as seen in Eq.~\eqref{e:Kbreve}. The trade off is the quadratic curvature terms over the manifold $\mathcal{M}$ which we will show how to manage. Let us demonstrate these features by making the following expansions
\begin{subequations}\label{e:KK}
\begin{align}\label{e:KKRicci}
	\mathcal{K}_{\a\b}\mathcal{K}^{\a\b} =& \cK_{ab} \cK^{ab}~~~\\
	\label{e:KKRiemann}
	\cK^\a{}_{\b\m\n} \cK_\a{}^{\b\m\n} =& g_{ae} \cK^a{}_{bcd} \cK^{ebcd} - \lambda_0^2 K_{abc}K^{abc}~~~.
\end{align}
\end{subequations}
These expansions can be easily derived via use of the succinct form of the metric in Eq.~\eqref{e:BigGSuccinct} and taking into account that the only non-vanishing $\lambda$ components of either $\cK_{\a\b}$ or $\cK^\a{}_{\b\m\n}$ are $\cK^\l{}_{\b\m\n}$. Notice the expansion Eq.~\eqref{e:KKRiemann} contains the projective Cotton-York tensor, Eq. \eqref{e:CYT}, which we see leads to quadratic derivatives on $\mathcal{P}_{ab}$ in the action and thus  provides dynamics to the field equations for $\cP_{ab}$. Continuing our electromagnetic analogy, $K_{nab}$ is to $F_{ab}$ as $\cP_{ab}$ is to $A_a$. 

Expanding further the components of $\cK^a{}_{bcd}$, $\cK_{ab}$, and $\sqrt{|G|}$ from the previous section, the action can be written as
\begin{align}
	S_{TW} =& - \tfrac{1}{2 \tilde{\kappa}_0} \int d\ell \ell^{-1} \int d^\rd x \sqrt{|g|} \left(R + (\rd-1) \cP + 2 \Lambda_0 \right) - \tilde{J}_0 c \int d\ell \ell^{-1} S_{GB} \cr
	&+ \tilde{J}_0 c \int d\ell \ell^{-1} \int d^\rd x \sqrt{|g|} \left(\lambda_0^2 K_{abc} K^{abc} - \cP_{ab} \tilde{\cP}_*^{ab} - p(\rd)  \cP_{ab} \cP^{[ab]} \right)\\
\label{e:GB}
S_{GB}=& \int d^\rd x \sqrt{|g|}\left(R^2 -4R_{ab}R^{ab}+R\indices{^a _{bcd}}R\indices{_a ^{bcd}}\right)~~~,~~~p(\rd) = 2 (4 \rd^2 - 3 \rd -2 )
\end{align}
with the following definitions in a slightly different convention from \cite{Brensinger:2020mnx} 
\begin{align}
&\tilde{\mathcal{P}}_{ab}\equiv (\rd-1)\mathcal{P}_{ab}+2R_{ab}\\
&\tilde{\mathcal{P}}_*^{ab}\equiv (\rd-1)g^{ab}\tilde{\mathcal{P}}-2(2\rd-3)\tilde{\mathcal{P}}^{ab}~~~.
\end{align} 

Notice the term $\cP^{[ab]} = \cP^{ab} - \cP^{ba}$ in the action above. Generally, $\cP_{ab}$ and $R_{ab}$ are not independent as their antisymmetric parts are proportional to the curl of $\alpha_b$
\begin{align}
	\cP_{[ab]} = \tfrac{1}{\rd+1} R_{[ab]} = - \partial_{[a} \alpha_{b]}~~~.
\end{align}

Up until this point, the connection $\Gamma^a{}_{bc}$ has been incompatible with the metric $g_{ab}$. From here on in this paper, we set the connection $\Gamma^a{}_{bc}$ to be a Levi-Civita connection,
\begin{align}
        \G^{m}{}_{ab} = \frac{1}{2} g^{mn}(g_{n(a,b)} - g_{ab,n})~~~,
\end{align}
compatible with the metric $g_{ab}$. This forces $\alpha_a=-\tfrac{1}{\rd+1}\Gamma\indices{^e _{ea}}=-\tfrac{1}{\rd+1}\partial_a \ln \sqrt{|g|}=g_a$ and $\cP_{[ab]} = R_{[ab]} = 0$, reducing the action to
\begin{align}
\label{e:TW_action_Levi-Civita}
S_{TW}=&-\tfrac{1}{2\kappa_0} \int d^{\rd}x \sqrt{|g|}\big(R+(\rd -1)\mathcal{P}+2\Lambda_0\big)\nonumber\\
&+J_0 c\int d^{\rd}x \sqrt{|g|}\left[\lambda_0^2K_{abc}K^{abc}-\mathcal{P}_{ab}\tilde{\mathcal{P}}_*^{ab}\right]-J_0 c S_{GB}
\end{align}
where now we have from here on in this paper
\begin{align}
	K_{abc} = \nabla_{[b}\cP_{c]a}~~~,~~~\cP_{ab} = \cP_{ba}~~~,~~~R_{ab} = R_{ba}~~~.
\end{align}
We have performed the integrations over $\ell$, absorbing the result into a redefinition of the constants
\begin{subequations}\label{e:flint}
\begin{align}
        &\frac{1}{\tilde{\kappa}_0}\int_{\ell_i}^{\ell_f} d\ell \ell^{-1} = \frac{\ln(\ell_f/\ell_i)}{\tilde{\kappa}_0}\quad \Rightarrow \kappa_0 \equiv\frac{\tilde{\kappa_0}}{\ln(\ell_f/\ell_i)}\\
        &\tilde{J}_0\int_{\ell_i}^{\ell^f}d\ell \ell^{-1}=\tilde{J}_0\ln(\ell_f/\ell_i)\quad \Rightarrow J_0 \equiv \tilde{J}_0 \ln(\ell_f/\ell_i) \; .
\end{align}
\end{subequations}
This integration along the projective direction allows for a natural scaling of both the gravitational coupling constant $\kappa_0$ and projective angular momentum parameter $J_0$. Notice if $\ell_f$ and $\ell_i$ are chosen to grow the angular momentum parameter $\tilde{J}_0$ into a larger $J_0$, the gravitational constant $\tilde{\kappa}_0$ necessarily shrinks to the smaller $\kappa_0$. This could tie the weakness of the gravitational force to a large angular momentum of the Universe. The fact that the Gauss-Bonnet term is a topological invariant in four-dimensions removes any potential higher-order metric terms from the equations of motion. At this point it is also clear that the TW gravity action reduces to Einstein-Hilbert when $\mathcal{P}_{ab}=0$.



\section{Realization of NMC Inflation}\label{s:NMCInflation}
Inflation via non-minimal coupling has been investigated for at least three decades. In a paper published in 1990\cite{Fakir1990}, Fakir and Unruh sought to remedy the issue of generically large density perturbations inherent to chaotic inflation scenarios by removing the assumption of minimal coupling between the inflaton field and the Ricci scalar curvature. Indeed, inflation via this non-minimal coupling has been shown to result in acceptable values for the spectral index $n_s$ and tensor-to-scalar ratio $r$ \cite{Bezrukov2008,Park:2008hz,Pallis2010}. Additionally, NMC inflation has been shown to provide a natural mechanism for the reheating phase occurring after inflation \cite{Watanabe2007}. Initially, the form of the non-minimal coupling was commonly assumed to be $\phi^2 R$, while more recent studies have investigated coupling of the more general form $f(\phi)R$ \cite{Park:2008hz}. Inflationary scenarios based on the more general non-minimal coupling have since been shown to also result in viable experimental predictions for $n_s$ and $r$ \cite{Park:2008hz}. This section details the main result of this paper, which is how TW gravity naturally realizes inflation with a non-minimal coupling.

\subsection{Tensor decomposition of \texorpdfstring{$\mathcal{P}_{ab}$}{Pab}}
From here on out we write the characteristic projective length scale $\lambda_0$ in units of Planck length such that $\lambda_0=n_{\lambda} \sqrt{8 \pi} l_p = n_\lambda \sqrt{8 \pi G \hbar/c^3}$, where $n_\lambda$ is a dimensionless scaling. We also write the projective angular momentum parameter in units of $\hbar$ such that $J_0=n_J \hbar$, where $n_J$ is a dimensionless scaling. For the remainder of this paper, we take natural units $\hbar=c=1$. The only remaining units will be written in terms of the reduced Planck mass
\begin{align}
\label{e:planck_mass}
M_p=\sqrt{\frac{\hbar c}{8\pi G}}\; .
\end{align}
All physical constants will now be written as
\begin{align}\label{e:NattyUnits}
\kappa_0= n_\kappa M_p^{-2}, \quad \lambda_0 =n_\lambda M_p^{-1}, \quad J_0=n_J \hbar=n_J \; .
\end{align} 
Here we include arbitrary factors $n_\lambda$, $n_\kappa$, and $n_J$ noticing there is no naturalness argument to use to constrain $n_\lambda$ as it is the scale of the projective direction and we have no a priori notion of how large this scale should be. In the following, we seek to constrain $n_\lambda$ from experiment to give us a window into the size of the projective direction that gives rise to inflation. At the same time, a resulting constraint on $n_J$ gives us a window into the angular momentum scale involved in these projective directions as well. In a similar previous work~\cite{Brensinger:2020mnx}, we saw the angular momentum scale $J_0$ to be of the order of the observable Universe when the cosmological constant arising from the vacuum solution of TW gravity was used to constrain $J_0$. It is possible that a constraint on $n_\kappa$ and the rescaling of the parameters $\tilde{\kappa}_0$ and $\tilde{J}_0$ as shown in Eqs.~\eqref{e:flint} could be related to quantum gravitational effects. Further research into the quantization of this theory needs to be done to investigate this possibility.

Using the following tensor decomposition of $\mathcal{P}_{ab}$, we may cast the dynamics of TW gravity into its trace and traceless degrees of freedom
\begin{align}\label{e:PDecomp}
	\mathcal{P}_{ab} = \tfrac{M_p}{n_{\lambda}} \phi g_{ab} + w_0 W_{ab}~~~,~~~W = g^{ab} W_{ab} = 0~~~.
\end{align}
where the field dimensions are: $[\mathcal{P}_{ab}]=M^2$, $[\phi]=M$, $[g_{ab}]=M^0$, and $[W_{ab}]=M^{2}$. The constant $w_0$ is dimensionless while the factor of $(\lambda_0)^{-1}=M_p/n_\lambda$ is included on the $\phi$ term to provide the correct units for a scalar field $[\phi] = L^{-1} = M$ and to cancel the $\lambda_0^2$ proportionality factor on the kinetic term in Eq.~\eqref{e:TW_action_Levi-Civita} as shown below.  This leads to the following decomposition of the action,
\begin{align}
\label{e:STWdecomposed}
S_{TW}=&-\tfrac{M_p^2}{2n_\kappa}\int d^{\rd}x \sqrt{|g|}\left[f(\phi) R +2 \Lambda_0 \right] + w_0\int d^{\rd}x \sqrt{|g|} \mathcal{L}_W\cr
&+4 (\rd-1)n_J\int d^\rd x \sqrt{|g|}\left[\tfrac{1}{2}\nabla_a \phi \nabla^a \phi - V(\phi)\right] -J_0 c S_{GB} \\
\label{e:LWphi}
\mathcal{L}_W =& w_0 \tfrac{ 2 n_J n_\lambda^2}{M_p^2}\left[  \nabla^{m}W^{nb} \nabla_{[m}W_{n]b} + \tfrac{(\rd-1)(2 \rd -3)M_p^2}{n_\lambda^2}  W^{ab}W_{ab}\right] \hspace*{-2 pt} +4(2\rd-3)n_J  R_{ab}W^{ab}  \cr
&- \tfrac{4 n_J n_\lambda}{M_p} \nabla_a \phi \nabla_b W^{ab} - \tfrac{4 n_J n_\lambda}{M_p} W \left[  \square \phi + \tfrac{(\rd-1)M_p^3}{8 n_\l n_Jn_\kappa} f(\phi) + \tfrac{(\rd-1)M_p}{2n_\lambda}R\right]  + \hat{\lambda} W^2 \\
\label{e:fVnhat}
f(\phi) =& 1 + \tfrac{4 (\rd-2)(\rd-3)n_\kappa n_J}{n_\lambda M_{p}}\phi ~~~,~~~
V(\phi) = \tfrac{\rd M_p^4}{64 n_J ^2 n_\kappa^2 (\rd-2)(\rd-3)}(f(\phi)^2 - 1) ~~~.
\end{align}
The Lagrange multiplier\footnote{Technically, the Lagrange multiplier is $\hat{\lambda} + (\rd -1)^2 w_0 n_J$. In the classical theory, this can be reabsorbed with no loss of generality. In the quantum theory, this may have to be revisited.} $\hat{\lambda}$ is placed in by hand to enforce tracelessness of $W_{ab}$ at the equations of motion level but not at the Lagrangian level. This ensures that the equations of motion are the same whether performing the decomposition before or after they are derived from variation of the action. That is, varying the action Eq.~\eqref{e:TW_action_Levi-Civita} with respect to $g_{ab}, \mathcal{P}_{ab}$ and then performing the decomposition Eq.~\eqref{e:PDecomp} yields the same equations of motion as varying the action Eq.~\eqref{e:STWdecomposed} with respect to the set of fields $g_{ab}, W_{ab}, \phi$. 

We notice a potential $V(\phi)$ that is quadratic in $\phi$ has developed from the decomposition. This is as expected as the action Eq.~\eqref{e:TW_action_Levi-Civita} was quadratic in $\cP_{ab}$. The potential includes a linear term in $\phi$ which can be removed via  field redefinition $f(\phi) = \tilde{\phi}$ leading to a correction to the cosmological constant proportional to $M_p^2/n_J$. In fact, the opposite was investigated in~\cite{Brensinger:2020mnx} where a field redefinition was used to generate an Einstein-Hilbert term with cosmological constant, setting all other dynamical fields to zero. 

We notice that as the potential's dependence on $f(\phi)$ is purely second order the minimum of the potential occurs where $f(\phi_{\text{min}}) = 0$
\begin{align}
	V_{min} = -\tfrac{\rd M_p^4}{64 n_J^2 n_\kappa^2 (\rd-2)(\rd-3)}~~~\text{at}~~~\phi_{min} =& - \tfrac{n_\lambda M_p}{4 (\rd-2)(\rd-3)n_Jn_\kappa}~~~.
\end{align}
At this potential minimum, the Einstein-Hilbert term vanishes, as it is proportional to $f(\phi_{\text{min}})$ which vanishes, and the theory is that of the the rank-two, symmetric traceless field $W_{ab}$ coupled to the metric. An interesting future work would be to investigate the theory at and around this potential minimum and possible connections to the cosmological constant. Our focus in this paper will instead be on making connections to slow roll inflationary cosmology, through non-minimally coupled inflation, to which we turn in the next sections.

\subsection{Canonicalization of the scalar field}

We now consider Eq.~\eqref{e:STWdecomposed}, with the assumptions $w_0=0$ and $\Lambda_0=0$. in $\rd = 4$ dimensions. Taking $w_0=0$, we see $\mathcal{P}_{ab}$ has only a trace degree of freedom held by the scalar field $\phi$. The action in Eq.~\eqref{e:STWdecomposed} now reduces to 
\begin{align}
\label{e:NMC_Jordan_frame}
S=&-\tfrac{M_p^2}{2n_\kappa}\int d^4x \sqrt{|g|}f(\phi) R + 12 n_J \int d^\rd x \sqrt{|g|}\left[\tfrac{1}{2}\nabla_a \phi \nabla^a \phi - V(\phi)\right] 
\\
f(\phi) =& 1 + \tfrac{8}{\hat{n} M_{p}} \phi~~~,~~~V(\phi) = \tfrac{M_p^4}{32 n_J^2 n_\kappa^2}(f(\phi)^2 - 1) ~~~,~~~\hat{n}= \frac{n_\lambda}{n_J n_\kappa}
\end{align}
where we have neglected $S_{GB}$ as in $\rd=4$ dimensions the variation $\delta S_\text{GB}$ is a boundary term and thus adds nothing to the equations of motion~\cite{Lanczos:1938sf}. This action is seen to be a particular case of a non-minimally coupled inflaton action in Jordan-frame with a potential of the form $V(\phi)=A \phi^2+B\phi$.  We note that TW gravity has reduced in Eq.~\eqref{e:NMC_Jordan_frame} to a composite of the linear and quadratic potential cases studied by \cite{Park:2008hz}, \cite{kodama2021relaxing}, after setting their dimensionless coupling parameter $\xi = \frac{8}{\hat{n}}$ and our $n_\kappa = 1$.  Thus, our inflaton coupling parameter is formed from a combination of free parameters, as seen in Eq.~\eqref{e:NMC_Jordan_frame}, rather than being an additional free parameter of the model.   Furthermore, TW gravity as reduced to Eq.~\eqref{e:NMC_Jordan_frame} falls into the categorization of models in \cite{kodama2021relaxing} as an $\mathcal{F}$- dominant case.

To manipulate this action into a form where we can easily apply the usual slow-roll analysis, we transform from Jordan to Einstein frame via the conformal transformation
\begin{align}
\label{e:conformal_trans}
g_{ab}=e^{-2\omega}\tilde{g}_{ab}\;,
\end{align}
where $g_{ab}$ and $\tilde{g}_{ab}$ are Jordan and Einstein frame metrics, respectively. We demonstrate the details in switching from Jordan to Einstein frame in Appendix~\ref{a:conf}, which follows closely~\cite{Dabrowski:2008kx,Kaiser:2010ps}. For $\rd =4$ we have
\begin{align}
	\omega = \ln \sqrt{\frac{f(\phi)}{n_\kappa} }~~~.
\end{align}
Under a conformal transformation with this $\omega$, the two parts of the Lagrangian in Eq.~\eqref{e:NMC_Jordan_frame} transform as
\begin{align}
	- \tfrac{M_p^2}{2 n_\kappa} \sqrt{|g|} f(\phi) R =  - \tfrac{M_p^2}{2} \sqrt{|\tilde{g}|} &\left[\tilde{R} + \tfrac{24}{\hat{n} M_p f(\phi)} \tilde{\square}\phi - \tfrac{9}{2} \left(\tfrac{8}{ \hat{n}M_pf(\phi)}\right)^2 \tilde{\nabla}^a \phi \tilde{\nabla}_a \phi \right] \\
	12 n_J \sqrt{|g|} \left[\tfrac{1}{2} \nabla^a \phi \nabla_a \phi - V(\phi)\right] =&	12 n_J \sqrt{|\tilde{g}|} \left[\tfrac{1}{2} \tfrac{n_\kappa}{f(\phi)} \tilde{\nabla}^a \phi \tilde{\nabla}_a \phi - \tfrac{M_p^4}{32 n_J^2 }(1 - f(\phi)^{-2})\right] .
\end{align}
Substituting into the action Eq.~\eqref{e:NMC_Jordan_frame} and combining like terms we find
\begin{align}
\label{e:Action_Einstein1Kory}
&S=\int d^4x \sqrt{|\tilde{g}|}\left( -\frac{M_p^2}{2}\tilde{R}+A(\phi)\frac{1}{2}\tilde{\nabla}_a \phi\tilde{\nabla}^a \phi+B(\phi)\frac{1}{2}\tilde{\square} \phi-\tilde{V}(\phi)\right)
\end{align}
with
\begin{align}
	A(\phi) =& \frac{288}{\hat{n}^2 f(\phi)^2} \left( 1 + \tfrac{\hat{n}n_{\lambda}}{24} f(\phi)\right)~~~,~~~B(\phi) = - \frac{24 M_p}{\hat{n}f(\phi)} \\
	\tilde{V}(\phi)\equiv &\frac{3M_p^4}{8n_J}\bigg(1-\frac{1}{f(\phi)^2}\bigg)= \frac{6 M_p^4  \phi(M_p \hat{n} +4   \phi)}{n_J(M_p \hat{n}+8  \phi)^2}~~~. \label{e:Vtilde}
\end{align}
Integrating by parts the $B(\phi)$ term and neglecting boundary terms yields
\begin{align}
S=\int d^4x \sqrt{|\tilde{g}|}\Bigg(-\frac{M_p^2}{2}\tilde{R}+C(\phi)\frac{1}{2}\tilde{\nabla}_a \phi\tilde{\nabla}^a \phi-\tilde{V}(\phi)\Bigg)
\end{align}
\begin{align}
C(\phi)\equiv  \frac{12}{\hat{n}^2 f(\phi)^2} [8 + \hat{n}n_\lambda f(\phi)] = \frac{96}{\hat{n}^2 f(\phi)^2} \left[ 1 + \frac{\hat{n}n_\lambda}{8} + \frac{n_\lambda}{M_p}\phi \right]~~~.
\end{align}
Defining the canonical field as~\footnote{Technically the canonical field $h$ would be defined as plus or minus that in Eq.~\eqref{e:hcandef}. This sign does not affect the analysis of the slow roll parameters, which ultimately only depend on $\phi$ and $h^2$, so we simply choose the plus solution hereafter for simplicity.}
\begin{align}\label{e:hcandef}
	\frac{dh}{d\phi} = \sqrt{C(\phi)}
\end{align}
\\
leads to the canonical scalar field action
\begin{align}\label{e:Canonical}
S=\int d^4x \sqrt{|\tilde{g}|}\Bigg(-\frac{M_p^2}{2}\tilde{R}+\frac{1}{2}\tilde{\nabla}_a h \tilde{\nabla}^a h-\tilde{V}(\phi(h))\Bigg)\;.
\end{align}
The differential equation Eq.~\eqref{e:hcandef} defining the canonical field $h$ can be solved piecewise exactly by separation of variables
\begin{align}\label{e:hphi}
\hspace*{-10 pt}
	h(\phi) =& \left\{ \begin{array}{ll}
	\sqrt{6} M_p \left[ \sqrt{1 + \tfrac{\hat{n}n_\lambda}{8} f(\phi)} - \tanh^{-1} \sqrt{1 + \tfrac{\hat{n}n_\lambda}{8} f(\phi)} \right] & -\tfrac{8}{\hat{n}n_\lambda} < f(\phi) < 0 \\ \\ 
	\sqrt{6} M_p \left[ \sqrt{1 + \tfrac{\hat{n}n_\lambda}{8} f(\phi)} - \coth^{-1} \sqrt{1 + \tfrac{\hat{n}n_\lambda}{8} f(\phi)} \right] &  f(\phi) >  0 	~~~.
	\end{array}
	\right.
\end{align}

The potential minimum $\tilde{V} = - \infty$ corresponds to the point $h = - \infty$.
 In the large  $\phi$ limit, this relationship becomes
\begin{align}
	h \approx \sqrt{6 M_p n_\lambda \phi} \left[ 1 - \left(1 - \tfrac{\hat{n} n_\lambda}{8} \right) \tfrac{M_p}{2 n_\lambda \phi} \right]
\end{align}
which can be inverted to solve for $\phi$
\begin{align}
	\phi \approx \frac{h^2}{6 M_p n_\lambda} \left(1 + \left(1 - \tfrac{\hat{n} n_\lambda}{8} \right) \tfrac{6 M_p^2}{h^2} \right)~~~.
\end{align}

\subsection{Slow-Roll Parameters, Observables, and the Corresponding Range of TW Gravity Parameters}
At this point, we can perform the calculation of slow-roll parameters and their corresponding observables in terms of the canonicalized scalar field $h$ in the Einstein frame. 
A general review of slow-roll parameters and scalar field inflationary cosmology is given in appendix~\ref{a:CosmoReview}.  
We first calculate the slow-roll parameters $\epsilon$ and $\eta$ using the equations for a canonical scalar field
\begin{align}
\epsilon=\frac{M_p^2}{2}\left(\frac{\partial\tilde{V}/\partial h}{\tilde{V}}\right)^2, \quad \eta=M_p^2\frac{\partial^2\tilde{V}/\partial h^2}{\tilde{V}} ~~~.
\end{align}

As we don't have a closed form solution for $\tilde{V}$ in terms of $h$, we cast these in terms of derivatives of $\phi$ via use of the chain rule and the relationship in Eq.~\eqref{e:hcandef}
\begin{align}
\epsilon=\frac{M_p^2}{2 C}\left(\frac{\partial \tilde{V}/\partial \phi}{\tilde{V}}\right)^2~~~,~~~\eta = \frac{M_p^2}{\sqrt{C} \tilde{V}} \frac{\partial}{\partial \phi}\left[C^{-1/2} \frac{\partial \tilde V}{\partial \phi} \right]~~~.
\end{align}
Carrying out the derivatives and simplifying yields the following forms of the slow-roll parameters
\begin{align} \label{e:epsilonetaf}
	\epsilon =& \frac{32}{3 (f^2 - 1)^2(8 + \hat{n} n_\l f)}~~~,~~~\eta = - \frac{16}{3} \frac{32+5 \hat{n} n_\l f }{(f^2 - 1) \left( 8 +\hat{n} n_\l f\right)^2}~~~.
\end{align}
which, in the large field limit become
\begin{align}\label{e:slow_roll_region}
\epsilon \simeq \frac{M_p^5 \hat{n}^4}{3072 n_\l  \phi^5}, \quad \eta\simeq -\frac{5M_p^3 \hat{n}^2}{96 n_\l   \phi^3} ~~.
\end{align}

We see that the necessary conditions $\epsilon \ll 1$ and $|\eta| \ll 1$, for the slow-roll approximation to hold, are satisfied in the large field limit of Eq.~\eqref{e:slow_roll_region}. Once we have solved for these slow-roll parameters, we will use them to calculate the scalar-mode spectral index $n_s$, scalar-mode amplitude $A_s$, and tensor-to-scalar amplitude ratio $r$ via the equations
\begin{align}
\label{e:ns_r_slow_roll}
n_s=1-6\epsilon +2\eta|_{\phi_{*}}\;, \quad A_s= \frac{1}{24 \pi^2 M_p^4}\frac{\tilde{V}}{\epsilon}|_{\phi_*}\;,\quad r=16\epsilon|_{\phi_{*}}\;, 
\end{align}
where $\phi_*$ is the field-value corresponding to the number of \emph{e}-foldings during inflation. 
In order to evaluate these expressions we need to obtain $\phi_{*}$ as a function of \emph{e}-foldings $N$. We calculate this using the standard slow-roll expression 
\begin{align}
\label{e:Num_efolds}
N \simeq \frac{1}{M_p^2} \int_{h_{\text{end}}}^{h_*}\frac{\tilde{V}}{\partial \tilde{V}/\partial h} dh
\end{align}
as we do not have closed form expressions for $\tilde{V}$ in terms of $h$, we must change variables back to $\phi$ using $dh = \sqrt{C(\phi)} d\phi$ from Eq.~\eqref{e:hcandef}
\begin{align}
	N \simeq \frac{1}{M_p^2} \int_{\phi_{\text{end}}}^{\phi_*}\frac{\tilde{V} C}{\partial \tilde{V}/\partial \phi} d\phi~~~.
\end{align}
Integrating this equation yields
\begin{align}\label{e:Nfun}
	N \simeq& \tfrac{1}{32} \left(  \hat{n}n_\l (\a^3-1) f_\text{end}^3+12 (\alpha^2 - 1) f_\text{end}^2   - 3 \hat{n}n_\l (\a-1) f_\text{end} - 24 \ln \a \right) \\
	\a \equiv& \tfrac{f(\phi_{*})}{f(\phi_\text{end})}~~~,~~~f_\text{end} = f(\phi_\text{end})~~~.
\end{align}
Since $\epsilon_{\phi_{\text{end}}} \simeq 1$ signals the approximate end of inflation, we can solve Eq.~\eqref{e:epsilonetaf} for the product $\hat{n}n_{\lambda}$ as a function of $f_\text{end}$
\begin{align}\label{e:nhnl}
    \hat{n}n_\lambda = \frac{32}{3 f_\text{end} (f_\text{end}^2 - 1)^2} - \frac{8}{f_\text{end}}~~~.
\end{align}

TW gravity has three free parameters ($n_J$, $n_\lambda$, $n_\kappa$) that we will fit using data from the Planck, BICEP2, and Keck array~\cite{PCX2019}. Our solution proceeds as follows
\begin{enumerate}
    \item Solve Eq.~\eqref{e:nhnl} for the range of $f_\text{end}$ corresponding to positive $\hat{n}n_{\lambda}$. Recall $\hat{n}$ was defined in terms of the three free parameters in Eq.~\eqref{e:fVnhat} which are all positive.
    \item Plug Eq.~\eqref{e:nhnl} into Eq.~\eqref{e:Nfun}, and solve for $f_\text{end}$ as a function of $f_*$ for various efoldings $N$. 
    \item For each of these solutions for $f_*$ and corresponding $\hat{n}n_{\lambda}$, calculate $r$ and $n_s$ from Eq.~\eqref{e:ns_r_slow_roll}.
    \item Fit these solutions to the Planck, BICEP2, and Keck array data~\cite{PCX2019}. For the range of solutions for $f_*$ that fits the $r$ and $n_s$ data, solve Eq.~\eqref{e:ns_r_slow_roll} for the corresponding range for $n_J$ that also fits the $A_s$ data.
    \item This will give a range of values for the three parameters ($n_J$, $n_\lambda$, $n_\kappa$) of the TW gravity model of slow roll inflation that fits the current inflatiory data~\cite{PCX2019}.
\end{enumerate}

\subsubsection{Constraining \texorpdfstring{$f_*$}{f*} and \texorpdfstring{$f_\text{end}$}{fend} for Various Efolds}\label{s:Constrainingf}
\begin{figure}[H] 
    \centering
    \figtitle{Constraining $f_\text{end}$ with $\hat{n}n_\lambda$}
	\includegraphics[scale=.75]{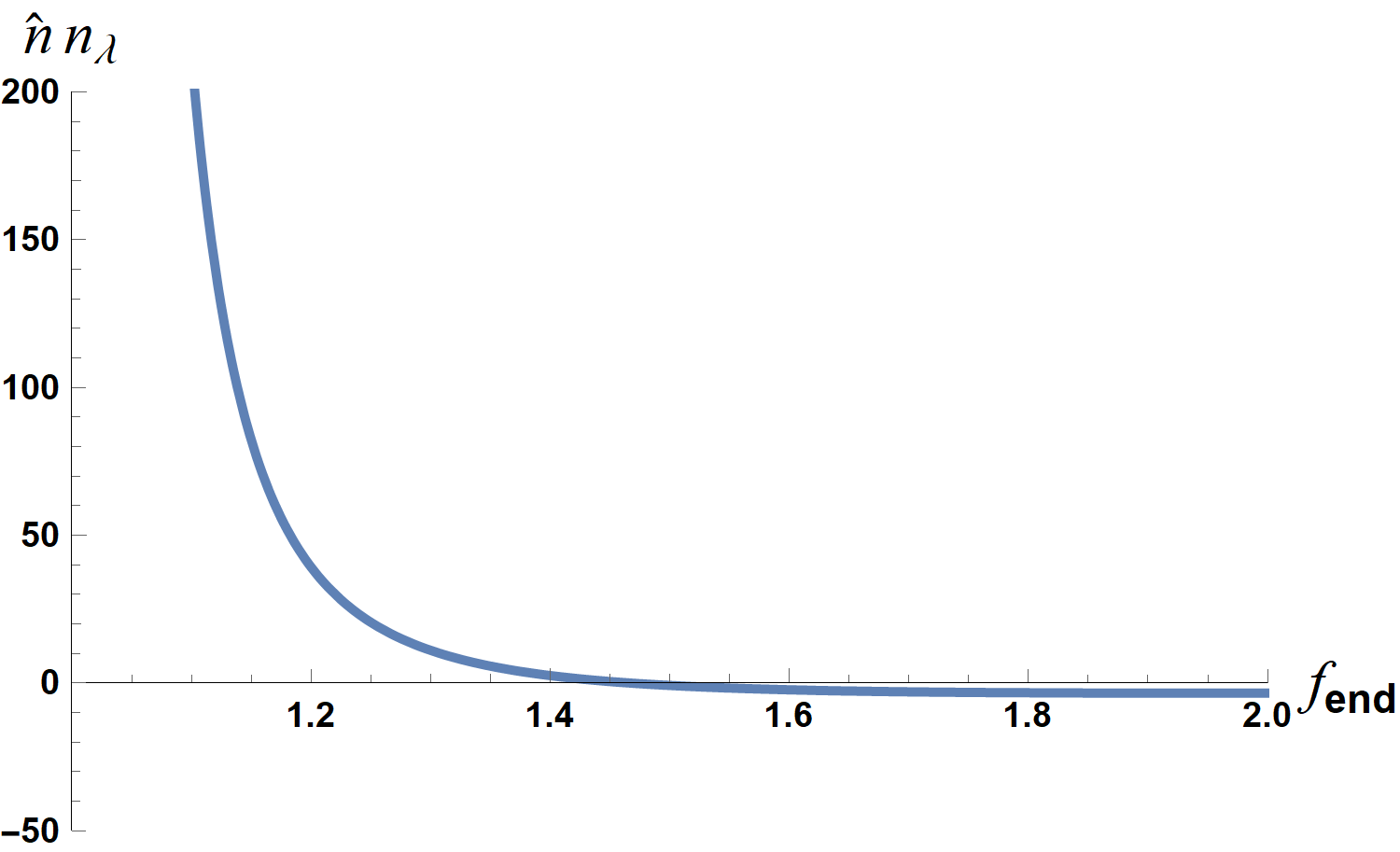}
	\caption{Solution to Eq.~\eqref{e:epsilonetaf} when $\epsilon_{\phi_{\text{end}}} = 1$.  The vertical asymptote is at precisely $f_\text{end} = 1$ and the $y$-axis crossing occurs at $f_\text{end}=\sqrt{1+2/\sqrt{3}} \approx 1.47$. }
	\label{fig:nahtnlamf}
\end{figure}

Figure \ref{fig:nahtnlamf} demonstrates that the valid range of solutions for Eq.~\eqref{e:nhnl} are for the range $1 < f_\text{end} < \sqrt{1 + 2/\sqrt{3}} \approx 1.47$. The lower limit  $f_\text{end} > 1$ can be seen to arise from Eq.~\eqref{e:Vtilde} where $\tilde{V}$ must be positive to ensure the slow roll approximation $\tilde{V} >> \dot{h}^2/2$ is valid. An  asymptote $\hat{n} n_\lambda \to \infty$ occurs at this lower limit. The upper limit $f_\text{end} <  \sqrt{1 + 2/\sqrt{3}}$ is enforced by the fact that $\hat{n} n_\lambda$ must be positive (the free TW gravity parameters all must be positive). 

\begin{figure}[H] 
    \centering
    \figtitle{Non-minimal coupling $f(\phi)$}
	\includegraphics[scale=.5]{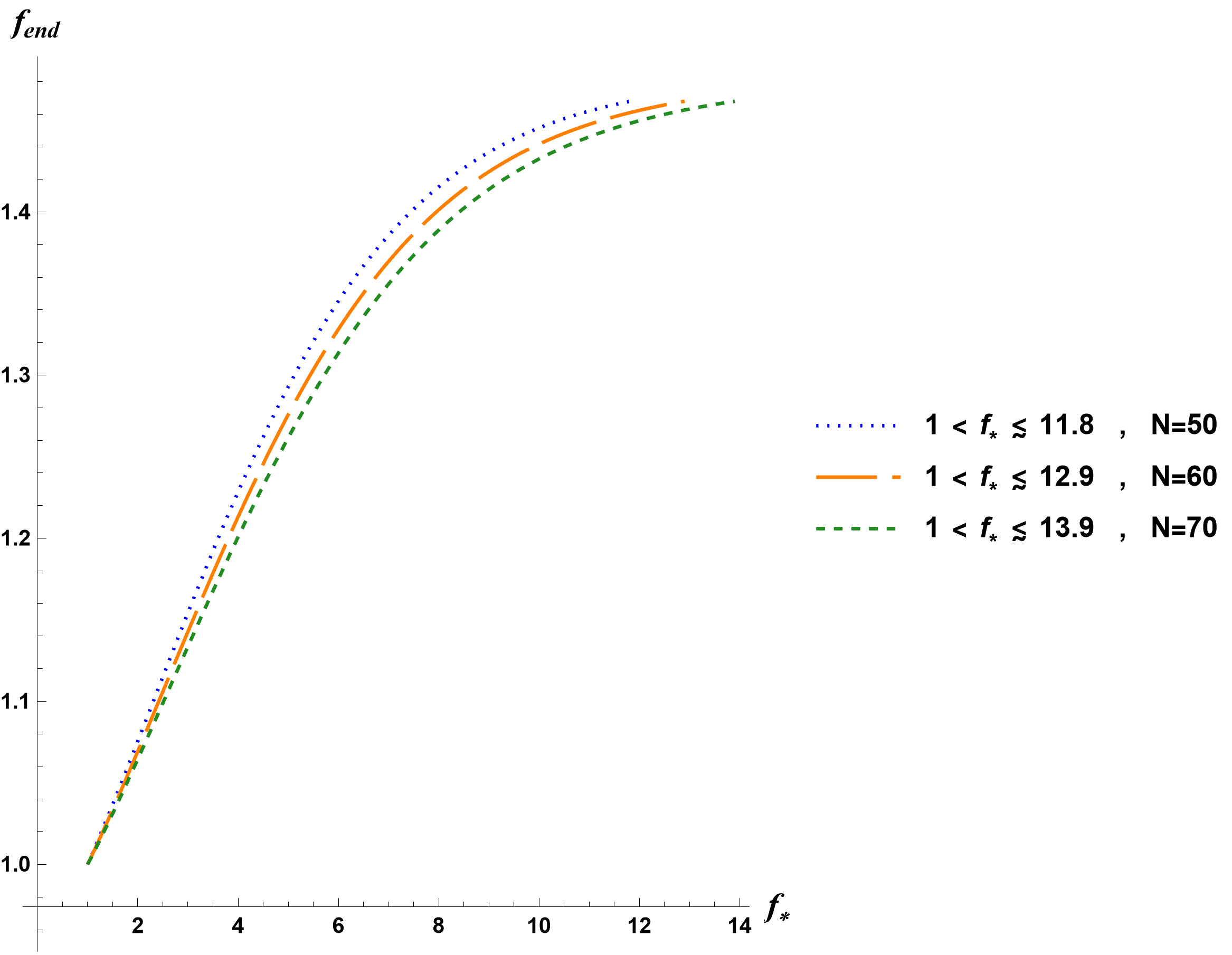}
	\caption{Solutions to Eqs.~\eqref{e:Nfun} and~\eqref{e:nhnl} for $N$ values of 50, 60, and 70. All satisfy $f_* > f_\text{end} > 1$ that is a necessary condition for inflation in the slow roll approximation.} 
	\label{fig:fstartfend}
\end{figure}

Next, we insert this range of solutions for $\hat{n}n_\lambda$ into Eq.~\eqref{e:Nfun} and solve for $f_\text{end}$ as a function of the initial condition $f_* = f(\phi_*)$. Solutions are plotted in Figure ~\ref{fig:fstartfend} for the number of efolds $N=50$, $60$, and $70$. These solutions all satisfy the necessary condition $f_* > f_\text{end}$ which ensures that inflation occurs prior to the condition $\epsilon_{\phi_*} \approx 1$ is realized, which shuts off inflation. These solutions also satisfy $f_\text{end}>1$ necessary for the validity of the slow roll approximation as previously described. 

The range of $f_*$ corresponding to positive $\hat{n}n_\lambda$ is to three significant figures as follows
\begin{subequations}
\label{e:fstarRange}
\begin{align}
    &1.00 < f_* \lesssim 11.8~~~N=50 \\ 
    &1.00 < f_* \lesssim 12.9~~~N=60 \\ 
    &1.00 < f_* \lesssim 13.9~~~N=70 
\end{align}
\end{subequations}
 Each upper limit on $f_*$ results in the upper limit for $f_\text{end} = \sqrt{1 + 2/\sqrt{3}}$ for which $\hat{n} n_\lambda \to 0$ as previously explained.

\subsubsection{Fitting the Planck, BICEP2, and Keck Array Data}
\begin{figure}[H] 
    \centering
    \figtitle{Observational Comparison (Plank+BICEP2/Keck array)}
	\includegraphics[scale=.5]{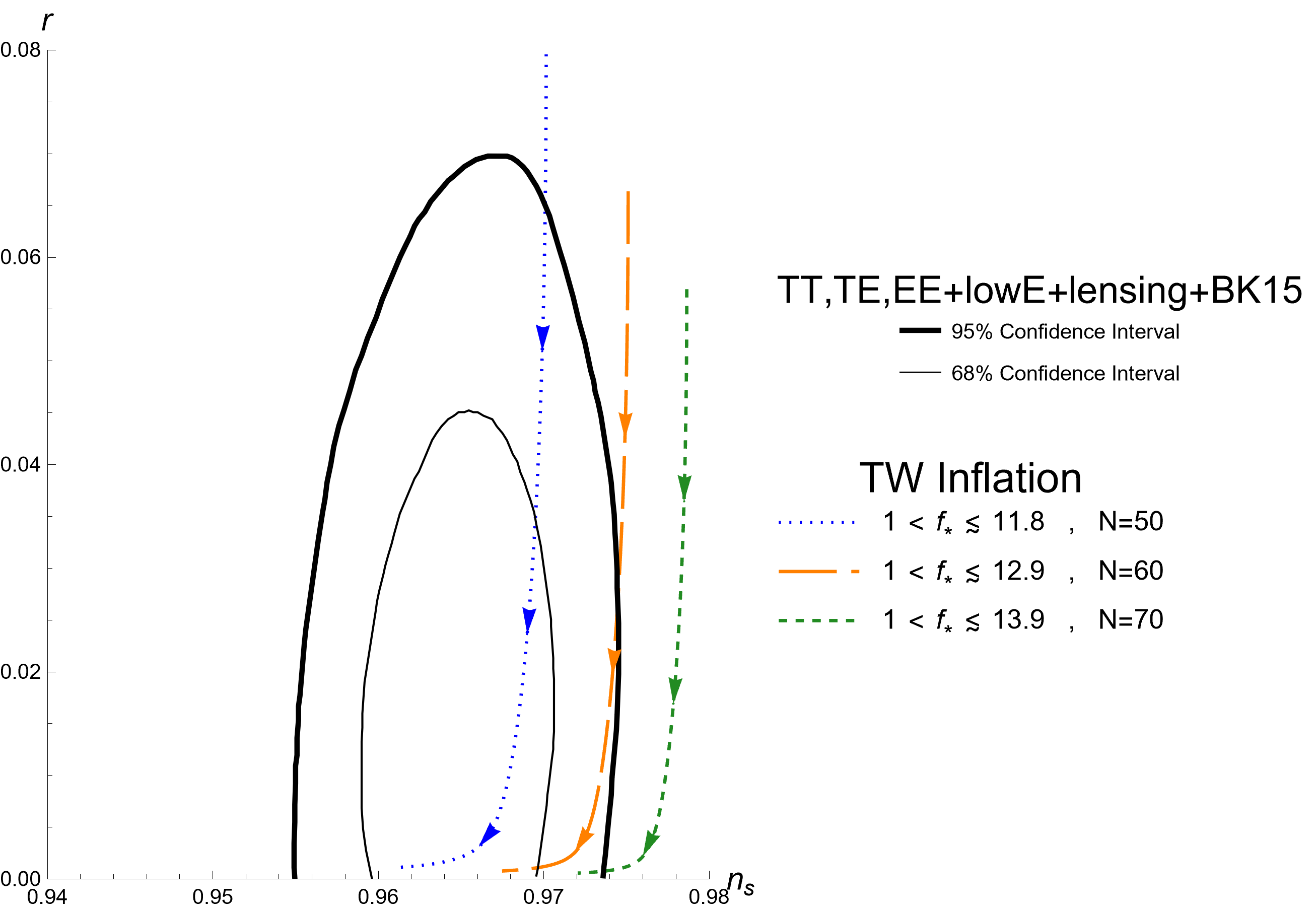}
	\caption{The tensor-to-scalar ratio $r$ and spectral index $n_s$ of the TW gravity slow roll inflation model are plotted over the valid range of $f_*$ in Eq.~\eqref{e:fstarRange}.  The three arrows on each line are located at $f_* = 1.30$, 2.00, and 6.00 to show the direction of increasing $f_*$ values.}
	\label{fig:ns_vsra}
\end{figure}

Now that we have valid ranges for $f_*$ for various efolds $N$, using Eq.~\eqref{e:ns_r_slow_roll} we can calculate a range of $n_s$ and $r$ predicted by our TW gravitational model. Using the most recent combination of Planck, BICEP2, and Keck array data presented in \cite{PCX2019} to constrain $n_s$ and $r$, we are able to place constraints on the product of dimensionless constants $\hat{n}n_{\lambda}$. In Figure \ref{fig:ns_vsra} we display the 68$\%$ and $95\%$ confidence intervals from \cite{PCX2019} with predictions from our TW gravitational model overlayed. 
Figure \ref{fig:ns_vsra} is in agreement with the predictions of \cite{kodama2021relaxing} for the case of chaotic inflation with power-law $\mathcal{F}$ in the metric formulation for $(n,p)=(1,1)$.  

Critical values for the parameters that yield values for $r$ and $n_s$ at the confidence level boundaries of Figure ~\ref{fig:ns_vsra} are listed in Table \ref{tab:ranges}. We see that the range of $\hat{n}n_{\lambda}$ values that fits the currently accepted confidence intervals for $n_s$ and $r$ depends heavily on the number of efolds $N$.  Smaller values of $N$ have a larger range of values of $\hat{n}n_\lambda$ that fit within the confidence intervals than larger values of $N$.

\begin{table}[H]
    \centering
    \begin{tabular}{c|c|c|c|c|c|c|}
        \cline{2-7}
         &\multicolumn{3}{c|}{\textbf{95\% CL Boundary}} 
         &\multicolumn{3}{c|}{\textbf{68\% CL Boundary}} \\
        \hline
         \multicolumn{1}{|c|}{$\boldsymbol{N}$} & $\boldsymbol{f_*}$ & $\boldsymbol{f_\text{end}}$ & $\boldsymbol{\hat{n}n_{\lambda}}$
        &  $\boldsymbol{f_*}$ & $\boldsymbol{f_\text{end}}$ & $\boldsymbol{\hat{n}n_{\lambda}}$ \\
        \hline
         \multicolumn{1}{|c|}{50} & 1.13 & 1.01 & 30300 & 1.62 & 1.05 & 1150  \\
        \hline
         \multicolumn{1}{|c|}{60} & 1.70 & 1.05 & 1060  & 9.84 & 1.44 & 0.896  \\
        \hline
        \multicolumn{1}{|c|}{70} & 11.1 & 1.45 & 0.625  \\
        \cline{1-4}
    \end{tabular}
    \caption{Parameters yielding  values of $r$ and $n_s$ at the 95\% and 68\% confidence level boundaries. Note that for $N=70$ there is no 68\% confidence level data as shown in Figure \ref{fig:ns_vsra}.}
    \label{tab:ranges}
\end{table}

\begin{figure}[H] 
    \centering
    \figtitle{$n_J$ Lower and Upper Limits (N=60)}
	\includegraphics[scale=.65]{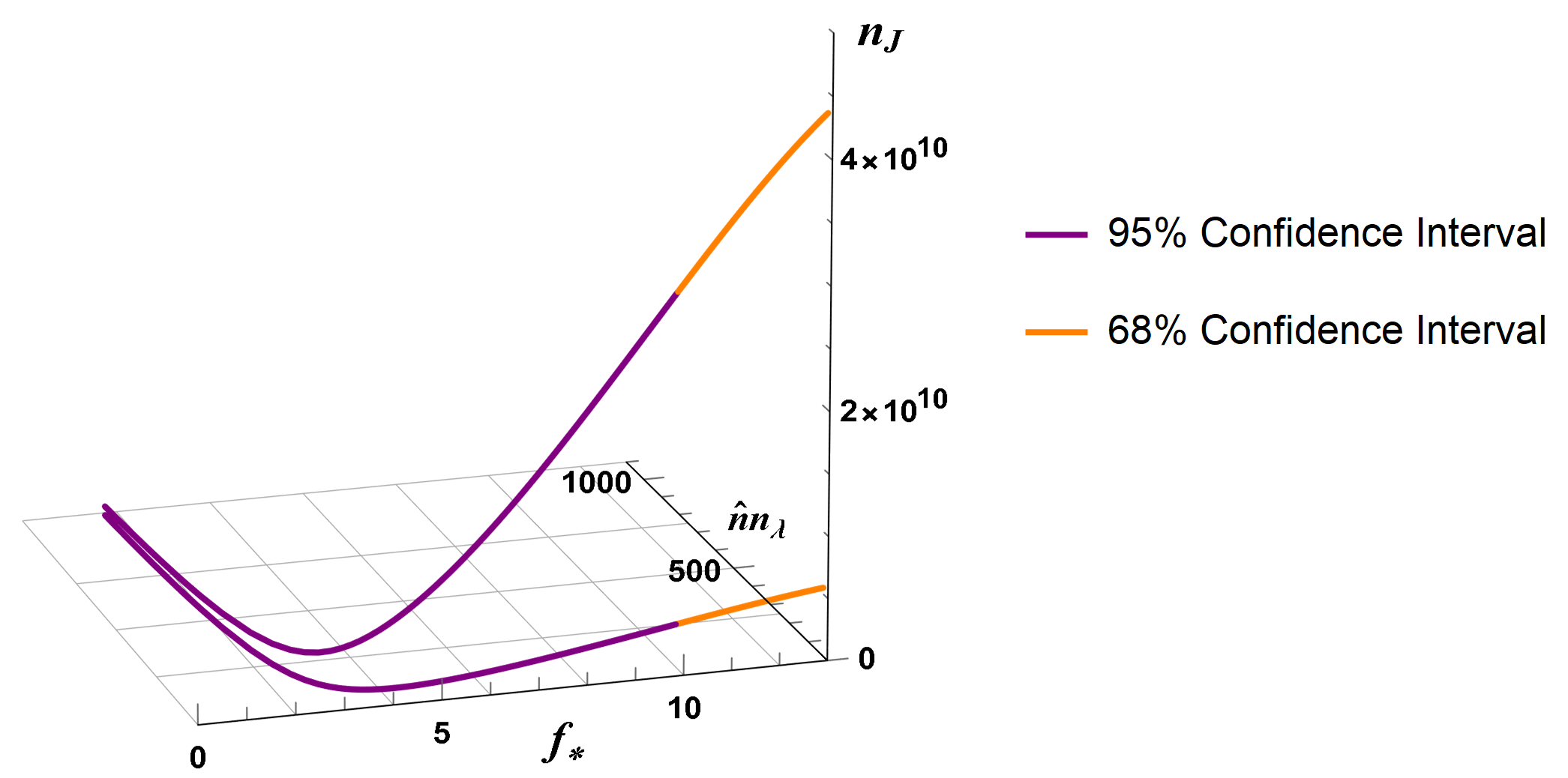}
	\caption{The lower and upper limits of $n_J$ are plotted versus $f_*$ and $\hat{n}n_{\lambda}$ for $N=60$ efolds. The corresponding confidence interval is also shown, using the boundary value of $f_*$ from Table \ref{tab:ranges}.  As $f_*$ decreases the range of $n_J$ decreases but does not vanish. For $N=60$ efolds, the range of values that fits completely within the 95\% confidence interval (including that part that is also within the 68\% confidence interval) is to three significant figures $1.15\times10^{8} \lesssim n_J \lesssim 4.37\times10^{10}$.}
	\label{fig:njrange}
\end{figure}

The observational constraints~\cite{PCX2019} for the scalar-mode amplitude $A_s$ are \begin{align}\label{e:AsConstraint}
    2 \le \ln 10^{10} A_s \le 4
\end{align}
Inserting Eq.~\eqref{e:ns_r_slow_roll} into Eq.~\eqref{e:AsConstraint}, for each bound we solve for the corresponding bounds on $n_J$. These bounds are shown in Fig.~\ref{fig:njrange} as a function of the solution of $f_*$ versus $\hat{n}n_\lambda$ that was extracted from Figs.~\ref{fig:nahtnlamf} and~\ref{fig:fstartfend} for $N=60$ efolds. In Fig.~\ref{fig:njrange} we see that $n_J \sim 10^{10}$ leads to the vast majority of the range of initial conditions $f_*$ consistent with the observational inflationary data plotted in Figure ~\ref{fig:ns_vsra}. 

An $n_J \sim 10^{10}$ points to an angular momentum scale $J_0 = n_J \hbar$ that is much less than angular momentum on cosmic scales such as the observable Universe. The constant $J_0$ would be on the order of the angular momentum of a few billion fundamental particles in the standard model with their spins aligned: roughly the number of electrons, protons, and neutrons in a SARS-CoV-2 virion~\cite{Covid:2019}

\subsubsection{Cosmological Constant Considerations}
\begin{figure}[H] 
    \centering
    \figtitle{Potential that Fits the 68\% Confidence Level Boundary Data}
	\includegraphics[scale=.75]{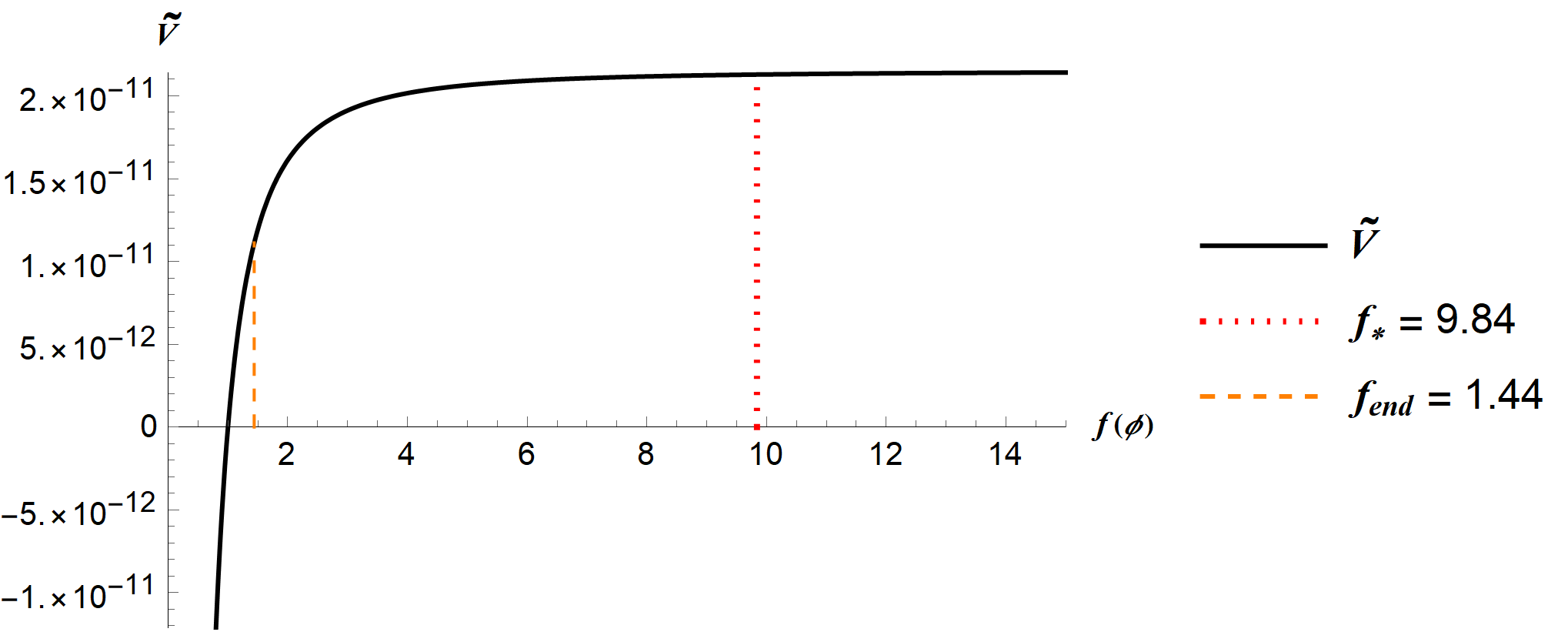}
	\caption{The potential in Eq.~\eqref{e:Vtilde} is plotted with respect to $f(\phi)$ for $n_J=1.74\times10^{10}$. Vertical lines indicate beginning and ending points of slow roll inflation that fit the 68\% confidence level boundary for $N=60$ efolds shown in Figure ~\ref{fig:ns_vsra}.  The potential $\tilde{V}$ is graphed in units of $\frac{M_pc^2}{(\sqrt{8\pi}l_p)^3}$, which is equivalent to $M_p^4$ in natural units.}
	\label{fig:potexp}
\end{figure}

In~\cite{Brensinger:2020mnx} it was shown how TW-gravity (in the $n_\kappa \to \infty$ limit of this paper) could give rise to corrections to the cosmological constant on the order of today's measured value, $\Lambda \sim 10^{-120} M_p^{2}$, if $n_J \sim 10^{120}$, corresponding to approximately the angular momentum scale for the observable universe.  In the general $n_\kappa$ case in Eq.~\eqref{e:Canonical}, setting the canonical scalar field $h=$~constant leads to contributions to the cosmological constant $\Lambda$ of the form
\begin{align}\label{e:CosmConst}
    \Lambda = M_p^{-2} \tilde{V}~~~.
\end{align}
The potential $\tilde{V}$ is plotted in Figure ~\ref{fig:potexp} for the case that fits the 68\% confidence level boundary for $N=60$ efolds as shown in Figure ~\ref{fig:ns_vsra}. 
The value $n_J = 1.74 \times 10^{10}$ has been chosen to lie within the range of values in Fig.~\ref{fig:njrange}. Note that changing the value of $n_J$ merely scales the vertical axis of Figure~\ref{fig:potexp} according to the inverse relationship in Eqs.~\eqref{e:Vtilde}. The range of the potential at the start, end, and throughout the entire epoch of inflation for $N=60$ which produces slow roll parameters that fit entirely within the 95\% confidence interval  is
\begin{subequations}
\begin{align}
    8.52 \times 10^{-{12}} M_p^4 &\lesssim \tilde{V} \lesssim 2.14 \times 10^{-9} M_p^4~~~\text{starting range} \\
    4.59 \times 10^{-{12}} M_p^4 &\lesssim \tilde{V} \lesssim 2.91 \times 10^{-{10}} M_p^4~~~\text{ending range} \\
    4.59 \times 10^{-{12}} M_p^4 &\lesssim \tilde{V} \lesssim  2.14 \times 10^{-9} M_p^4~~~\text{range throughout} 
\end{align}
\end{subequations}
where all lower bounds correspond to $f_\text{end} = \sqrt{1 + 2/\sqrt{3}}\approx 1.47~(f_* = 12.9)$, $n_J = 4.37 \times 10^{10}$ and all upper bounds correspond to $f_\text{end}  = 1.05 ~(f_* = 1.70)$, $n_J = 1.15 \times 10^{8}$ to three significant figures.

In Figure ~\ref{fig:potexp}, we have $\tilde{V}\sim 10^{-11} M_p^4$ at the end of inflation, thus predicting that $\Lambda \sim 10^{-11} M_p^{2}$ according to Eq.~\eqref{e:CosmConst}. This would be the case if the inflaton field is completely frozen at the end of inflation. The steepness of the slope at the end of inflation in Figure ~\ref{fig:potexp} suggests the possibility that $h$ is not completely frozen at the end of inflation and that $\tilde{V}$ might yet decrease substantially for a small change in $h$. Thus perhaps the prediction of $\Lambda \sim 10^{-11} M_p^{2}$ at the end of inflation is not the end of the story for TW-gravity born inflatons, but $\Lambda$ is actually much smaller.


\subsubsection{Validating the Slow Roll Approximation and LHC Considerations}
\begin{figure}[H]
     \centering
     \figtitle{Validating Slow-Roll, Maximum $f_* = 12.9$ for $N=60$}
     \begin{subfigure}[b]{0.496\textwidth}
         \centering
         \includegraphics[scale=.41]{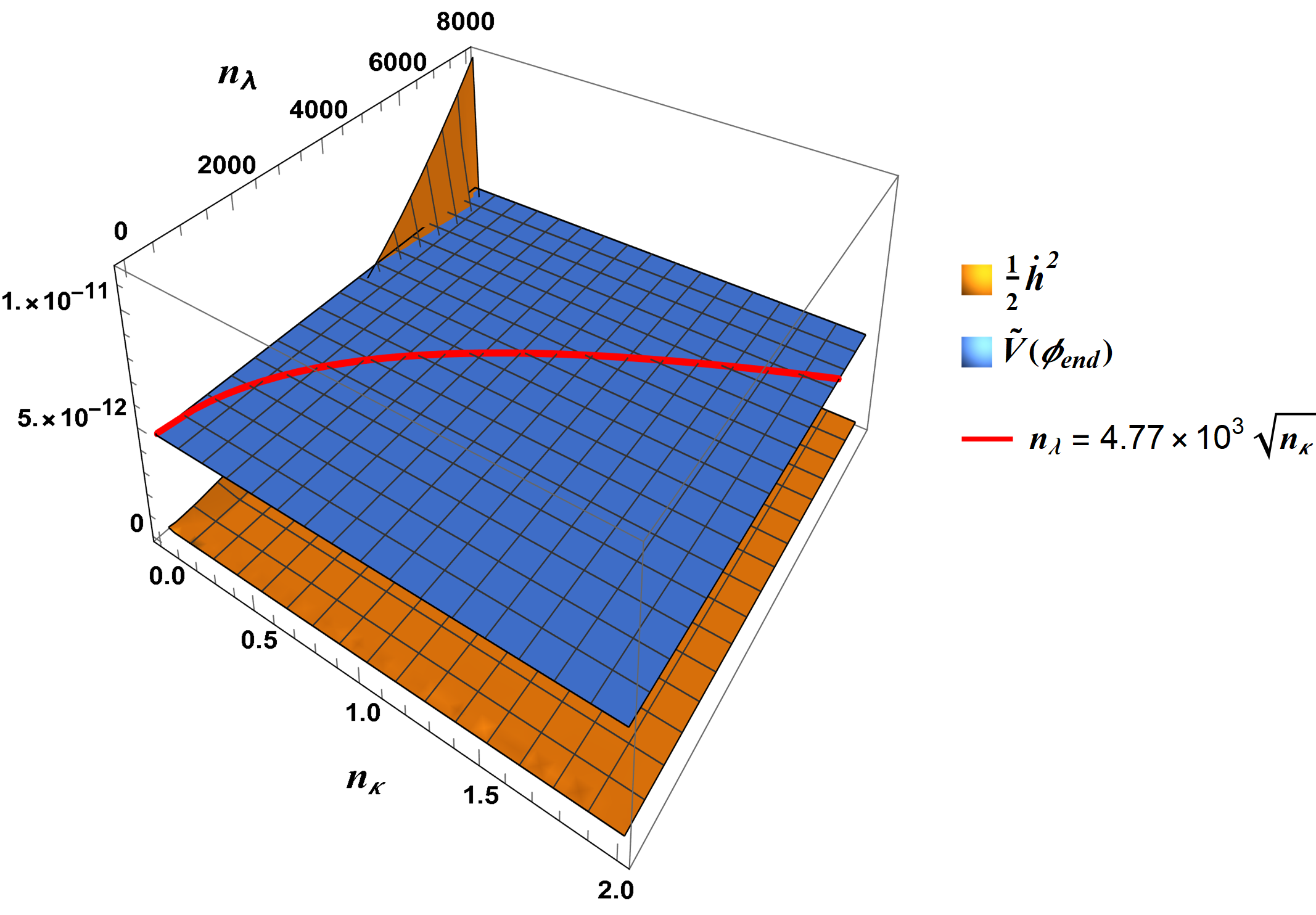}
         \caption{$n_J=4.37\times10^{10}$}
         \label{fig:12njup}
     \end{subfigure}
     \hfill
     \begin{subfigure}[b]{0.496\textwidth}
         \centering
         \includegraphics[scale=.41]{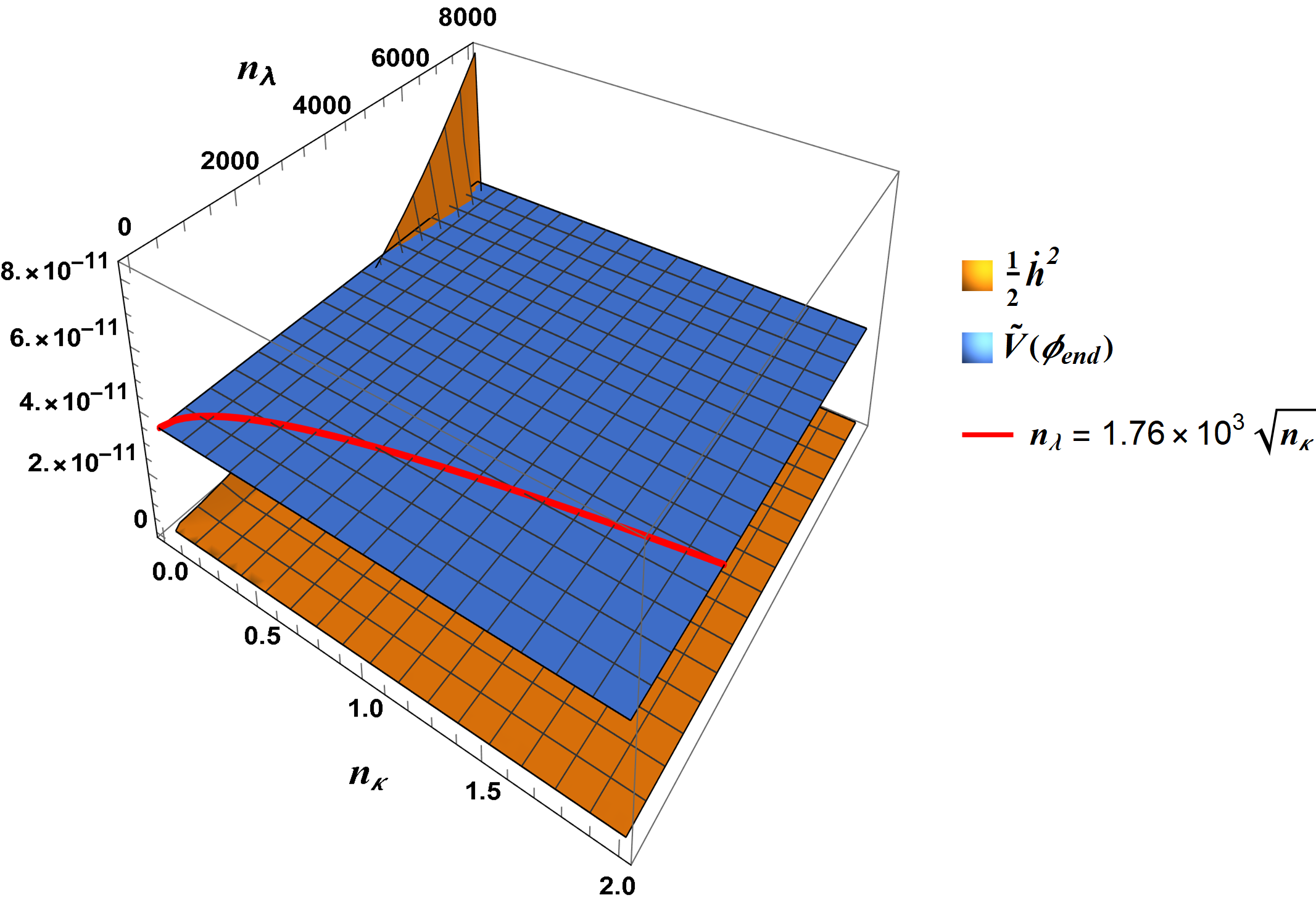}
         \caption{$n_J=5.92\times10^{9}$}
         \label{fig:12njlo}
     \end{subfigure}
    \caption{Kinetic energy $\tfrac{1}{2}\dot{h}^2$ and constant potential $\tilde{V}(\phi_\text{end})$ versus $n_\lambda$ and $n_\kappa$ for maximum $f_* = 12.9$ and $f_\text{end} = \sqrt{1 + 2/\sqrt{3}}$ for (a) upper bound on $n_J$ and (b) lower bound on $n_J$. The red line marks the solution $n_\lambda$ = $n_\lambda(n_\kappa)$ for fixed $n_J$.}
    \label{fig:fs12}
\end{figure}

Assuming inflation occurs at time scale of $10^{-36}s$, we can approximate the time derivative squared of the canonical scalar field $h$ as $\dot{h}^2 \approx ((h_\text{end} - h_*)/10^{-36}s)^2$, with $h_\text{end}$ and $h_*$ calculated from Eq.~\eqref{e:hphi} evaluated at $f_\text{end}$ and $f_*$, respectively. In this approximation, the following 3D plots are of $\dot{h}^2/2$ versus $n_\lambda$ and $n_\kappa$ for the upper and lower bounds of $n_J$ associated with the $A_s$ constraint Eq.~\eqref{e:AsConstraint} at $N=60$ efolds. The potential $\tilde{V}(\phi_\text{end})$~=~constant for given $n_J$, as calculated from Eq.~\eqref{e:Vtilde}, is used to demonstrate the validity of the slow roll approximation as it is the lowest value of the potential over the slow roll inflationary epoch. These plots clearly show that the slow roll approximation is valid for $N=60$ as the solution $n_\lambda$ = $n_\lambda(n_\kappa)$ for fixed $n_J$, marked by the red line, is clearly in the range where $\tilde{V} >> \dot{h}^2/2$.  Figure~\ref{fig:fs12} uses the maximum values  $f_* = 12.9$ and $f_\text{end} = \sqrt{1 + 2/\sqrt{3}}$ for $N=60$ efolds as described in section~\ref{s:Constrainingf}. 

\begin{figure}[H]
     \centering
     \figtitle{Validating Slow-Roll, 68\% Confidence Level Boundary Data for $N=60$}
     \begin{subfigure}[b]{0.496\textwidth}
         \centering
         \includegraphics[scale=.41]{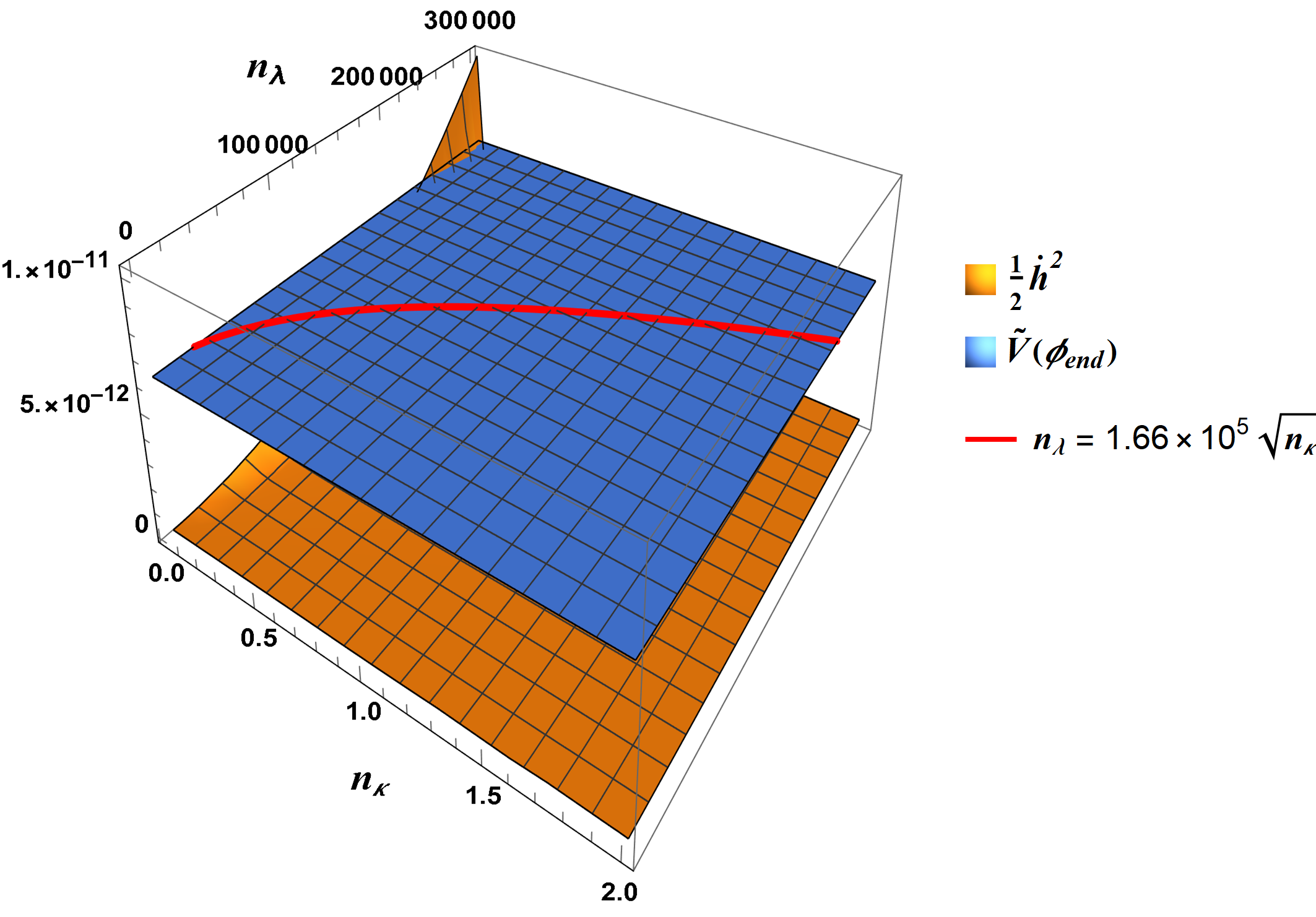}
         \caption{$n_J=3.07\times10^{10}$}
         \label{fig:9njup}
     \end{subfigure}
     \hfill
     \begin{subfigure}[b]{0.496\textwidth}
         \centering
         \includegraphics[scale=.41]{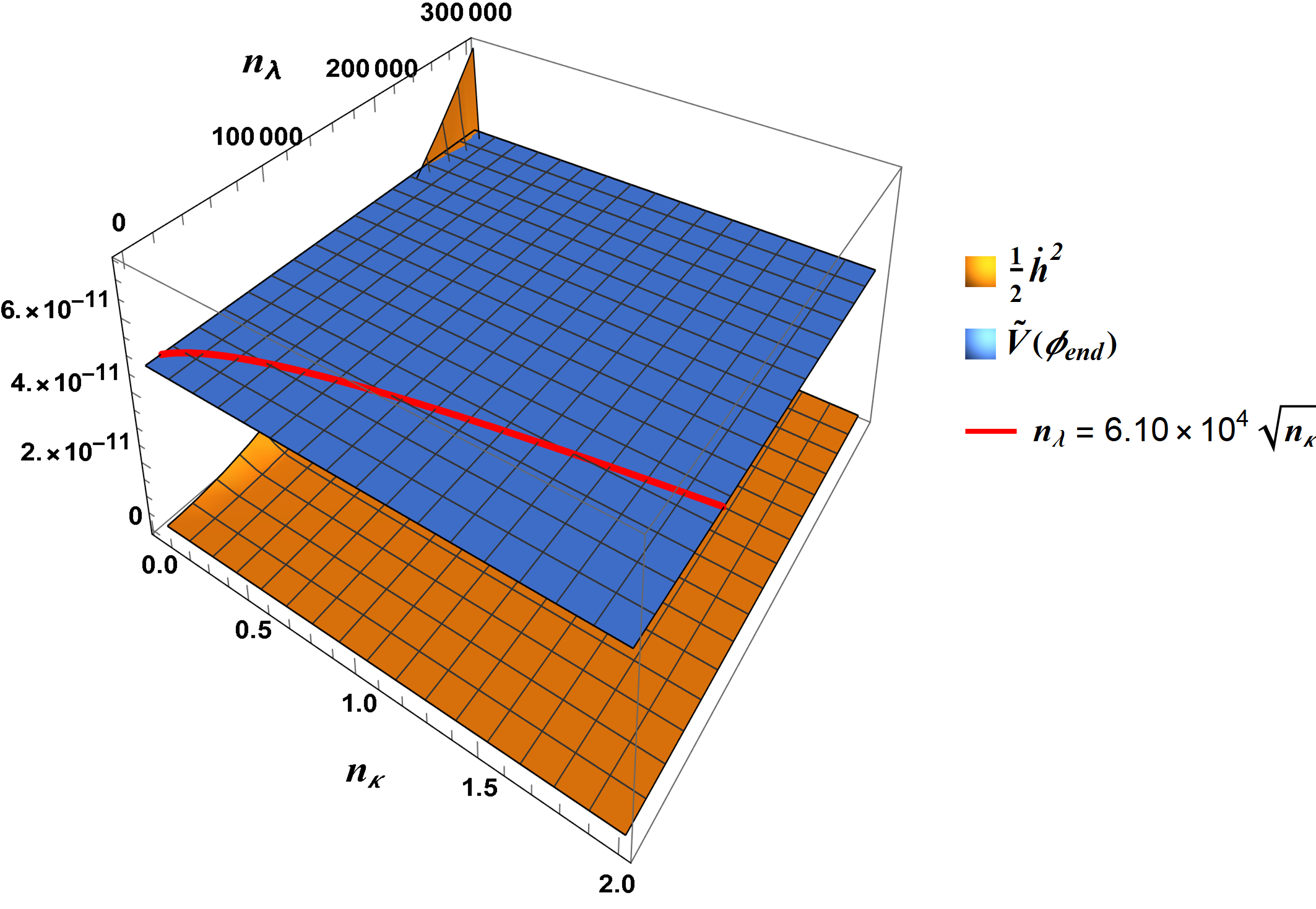}
         \caption{$n_J=4.16\times10^{9}$}
         \label{fig:9njlo}
     \end{subfigure}
    \caption{Kinetic energy $\tfrac{1}{2}\dot{h}^2$ and constant potential $\tilde{V}(\phi_\text{end})$ versus $n_\lambda$ and $n_\kappa$ for $f_* = 9.84$ and $f_\text{end} = 1.44$ for (a) upper bound on $n_J$ and (b) lower bound on $n_J$. The red line marks the solution $n_\lambda$ = $n_\lambda(n_\kappa)$ for fixed $n_J$. }
    \label{fig:fs9}
\end{figure}

\begin{figure}[H]
     \centering
     \figtitle{Validating Slow-Roll, 95\% Confidence Level Boundary Data for $N=60$}
     \begin{subfigure}[b]{0.496\textwidth}
         \centering
         \includegraphics[scale=.41]{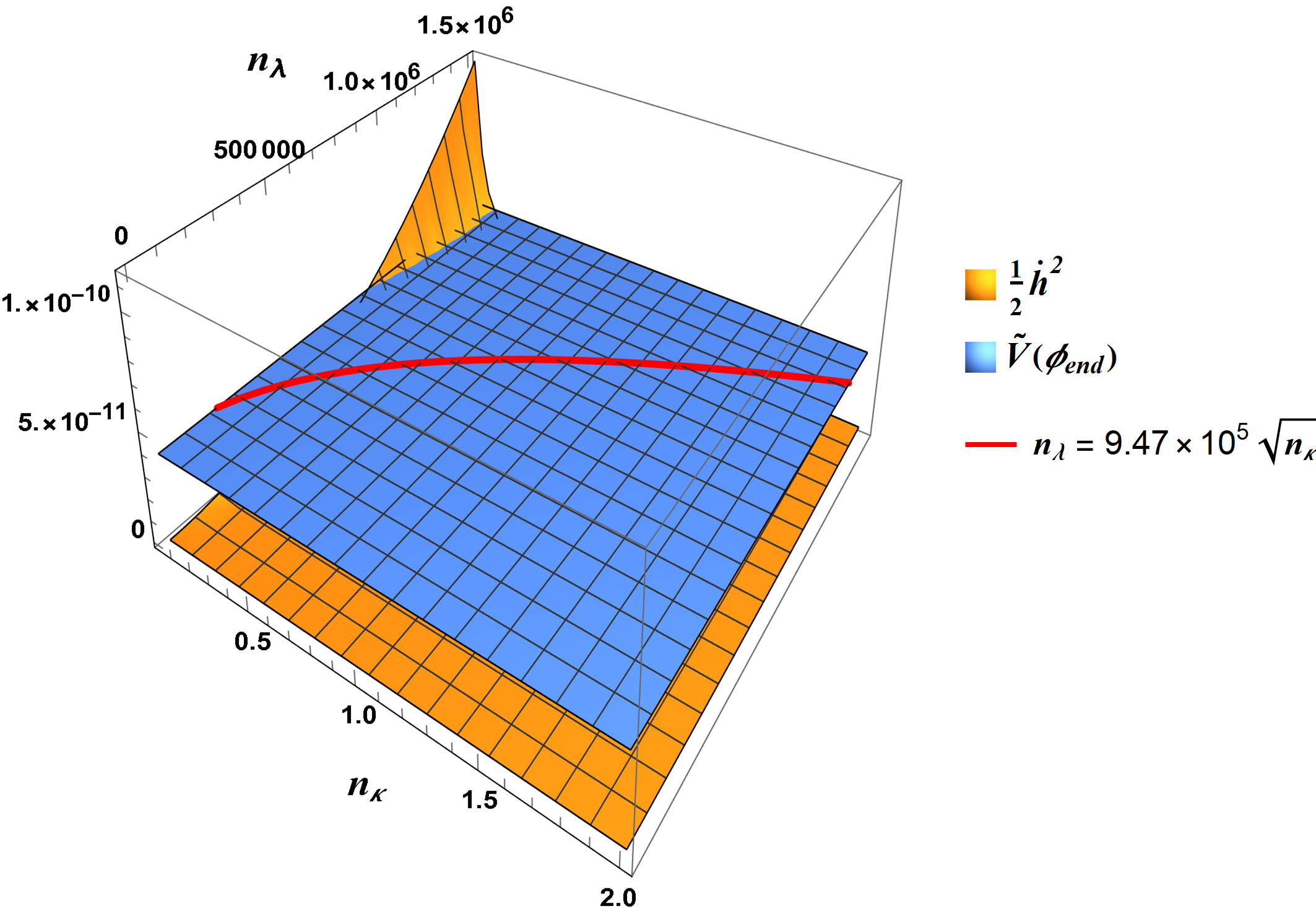}
         \caption{$n_J=8.47\times10^{8}$}
         \label{fig:1njup}
     \end{subfigure}
     \hfill
     \begin{subfigure}[b]{0.496\textwidth}
         \centering
         \includegraphics[scale=.41]{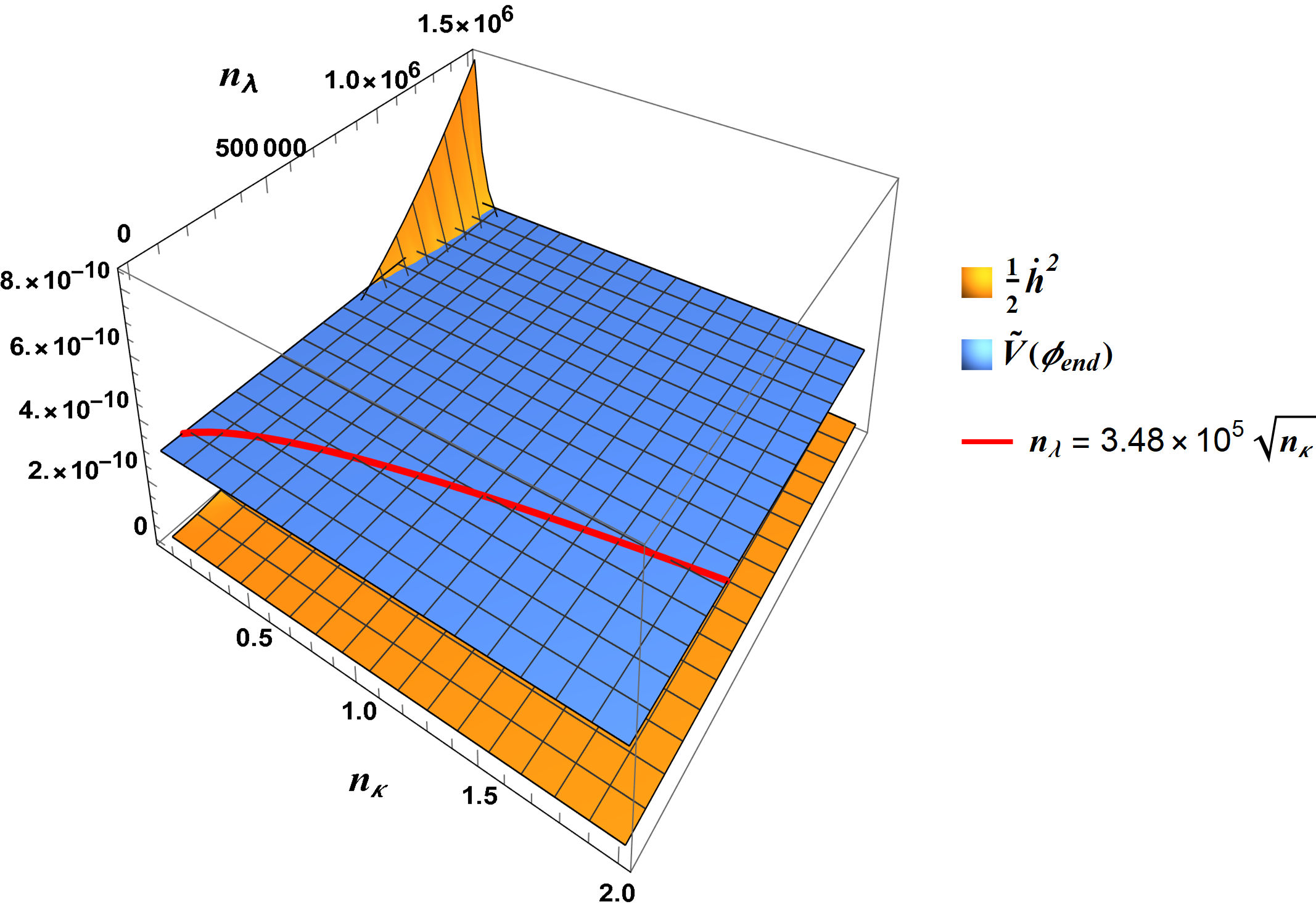}
         \caption{$n_J=1.15\times10^{8}$}
         \label{fig:1njlo}
     \end{subfigure}
    \caption{Kinetic energy $\tfrac{1}{2}\dot{h}^2$ and constant potential $\tilde{V}(\phi_\text{end})$ versus $n_\lambda$ and $n_\kappa$ for $f_* = 1.70$ and $f_\text{end} = 1.05$ for (a) upper bound on $n_J$ and (b) lower bound on $n_J$. The red line marks the solution $n_\lambda$ = $n_\lambda(n_\kappa)$ for fixed $n_J$. }
    \label{fig:fs1}
\end{figure}

Figures~\ref{fig:fs9} and~\ref{fig:fs1} use the values for $f_*$ and $f_\text{end}$ at the 65\% and 95\% confidence level boundaries given in Table~\ref{tab:ranges}, respectively. Plots for $N=50$ and $N=70$ similarly indicate the validity of the slow roll approximation for those values of \emph{e}-folds.

\begin{figure}[H]
            \centering
            \figtitle{LHC Sensitive Values for 95\% Confidence Level Boundary Data for $N=60$}
            \includegraphics[scale=.65]{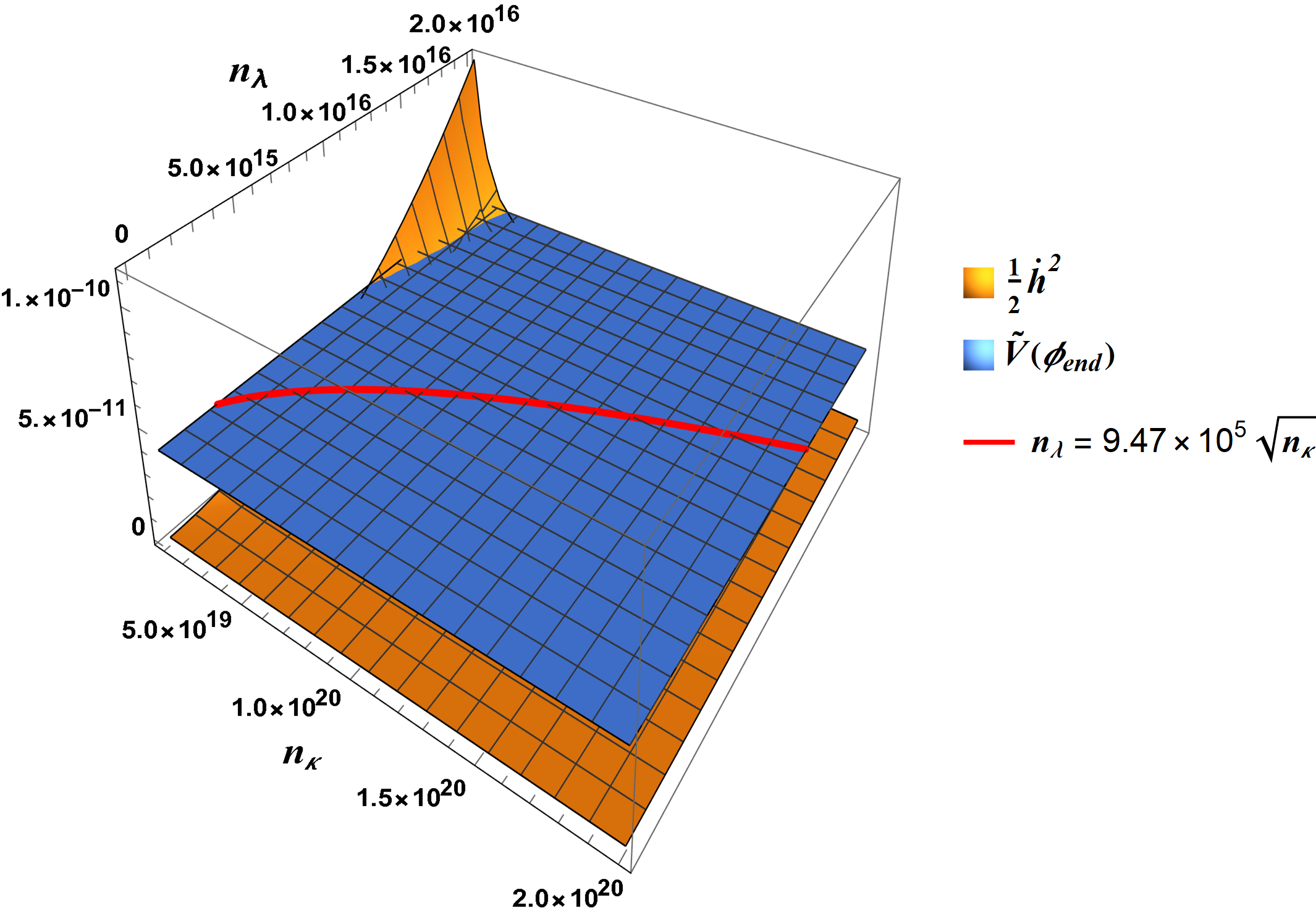}
            \caption{Kinetic energy $\tfrac{1}{2}\dot{h}^2$ and constant potential $\tilde{V}(\phi_\text{end})$ versus $n_\lambda$ and $n_\kappa$ for $f_* = 1.70$, $f_\text{end} = 1.05$, and $n_J=8.47\times10^{8}$. The red line marks the solution $n_\lambda$ = $n_\lambda(n_\kappa)$ for fixed $n_J$. This solution has precisely $n_{\kappa} = 1.10 \times 10^{20}$ for $n_{\lambda} = 10^{16}$. This is the smallest value of $n_{\kappa}$ that corresponds to an LHC-scale $n_{\lambda} = 10^{16}$, for all values of $f_*$ within the 95\% confidence level for $N=60$. This plot is the same as Fig.~\ref{fig:1njup} with a modified $n_\kappa$ and $n_\lambda$ range that is within scales that can be probed at the LHC. }
            \label{plt:LHC}
\end{figure}

Changing the scale of Figure~\ref{fig:1njup} as in Fig.~\ref{plt:LHC} allows for some analysis of LHC measurable effects. Figure~\ref{plt:LHC} demonstrates that $n_\kappa\sim 10^{20}$ for an $n_\lambda\sim 10^{16}$ that is associated with projective length scale effects that hypothetically could be probed at the LHC. As seen in Eq.~\eqref{e:NMC_Jordan_frame}, in the Jordan frame the gravitational coupling constant is proportional to $n_\kappa/(M_p^2 f(\phi))$. As $f(\phi)\sim 1$ over the range of solutions, this indicates that LHC measurable effects point to a very large gravitational coupling as viewed in the Jordan frame.

\section{Conclusion}
\label{s:conclusion}
We have shown that TW gravity provides a raison d'etre for inflation, the underlying foundational principle being projective symmetry. This symmetry itself arises from the dynamical extension of a coadjoint element of the Virasoro algebra to higher dimensions.  By decomposing the tensor field arising from TW gravity into trace and traceless degrees of freedom and setting the traceless components to zero, we have found that the resulting action reproduces that of (generalized) non-minimally coupled inflation with a novel inflaton potential.  After performing a conformal transformation to Einstein frame, we recover the canonical scalar field inflaton action, where the free parameters of TW gravity $n_J$, $n_\kappa$, and $n_\lambda$ (TW-parameters)  become embedded in the canonical field $h$ and its potential $\tilde{V}$. Calculating the slow roll parameters and applying conditions about the end of inflation allowed us to constrain relationships to the TW-parameters.  Finally, using recent observational data for the scalar-mode spectral index, scalar-mode amplitude, and tensor-to-scalar amplitude ratio, we determined the ranges for the TW-parameters that fit recent data for $N=50$, $60$, and $70$ efoldings. For $N=70$, however, the model does not fit within the most constraining confidence interval of the Planck+BICEP2/Keck array data. Generally, a lower number of efoldings paired with a larger value for the inflaton field at the start of inflation better matches the data.  Furthermore, we confirmed that the range of TW-parameters that fits the data is consistent with the slow-roll approximation of a potential dominated expansion. The range of parameters that fits within the 95\% confidence level is $1.15\times10^{8} \lesssim n_J \lesssim 4.37\times10^{10}$ and $0 < n_\lambda^2/(n_J n_\kappa) \lesssim 1030$ for $N=60$.

The TW-parameter $n_J$ corresponds to an angular momentum scale for TW gravity, associated with the TW coupling constant $J_0 = n_J \hbar$. We find $n_J \sim 10^{10}$ fits the slow roll cosmological data and leads to a cosmological constant contribution $\Lambda \lesssim 10^{-11} M_p^{2}$ from the scalar field settling down at the end of inflation. This correction is still much larger than the measured result $\Lambda \sim 10^{-120} M_p^{2}$, however, the shape of the TW-potential $\tilde{V}$ at the end of inflation indicates the possibility that $\Lambda$ could continue to decrease lower than $10^{-11} M_p^{2}$ after inflation was over. We would like to return to this in future works.

The inflationary solutions of TW gravity presented include a regime where a large Jordan frame gravitational coupling constant might lead to LHC sensitive TW effects.  Such effects may arise as dark matter portals through the spin connection of the higher-dimensional projective space~\cite{Brensinger:2017gtb,Brensinger:2020mnx,Brensinger:2020gcv}. Specifically, the spin connection gives rise to an axial coupling between the trace of $\mathcal{P}_{ab}$, that is the scalar field inflaton described here, and fermions~\cite{Brensinger:2020mnx,Brensinger:2020gcv}. The inflationary solutions of TW gravity indicate that if such a dark matter portal exists at LHC scales, it could shed light on inflationary cosmology as well. Furthermore, TW gravity will contribute to the reheating process after inflation, through direct decay of TW inflatons and possibly through facilitating decays as portals. Investigations in these directions will be pursued in the future.

In this paper, we assumed the connection was Levi-Civita. An interesting future work would be to relax this constraint and consider the Palatini approach of~\cite{Brensinger:2020gcv} that leads to a model with more tensorial degrees of freedom along with more equations of motion associated with the connection. We also plan to investigate how adding the traceless component of the tensor field $\mathcal{P}_{ab}$ in Eq.~\eqref{e:PDecomp} will affect the analysis.  Due to the addition of the Lagrangian Eq.~\eqref{e:LWphi}, we will need to solve the field equations directly and make an Ansatz for the form of the components of $W_{ab}$.  This could possibly provide a theoretical origin to the anisotropic seeds needed for galaxy formation in the early universe.


\section*{Acknowledgments}
The research of K.\ S.\ is supported in part by the endowment of the Ford Foundation Professorship of Physics at Brown University. The research of M. A., B. C., X. J., and M. H. K. were all supported by Summer Research Fellowships provided by Bates College. We thank Kenneth Heitritter, Vincent G. J. Rodgers, and Yehe Yan for helpful discussions.

\appendix

\section{Units and Conventions}
The units of the various constants used throughout this paper for $\rd = 4$ are
\begin{align}
\begin{split}
     [\phi] =& [h] = L^{-1}~~~,~~~  [\cP_{ab}] = [W_{ab}] = [\Lambda_0]= [R_{ab}] = L^{-2}~~~,~~~ [J_0] = \frac{M L^2}{T} ~~~, \cr
     [\lambda_0]   =& L ~~,~~ 
        [\ell] = [w_0] = \text{dimensionless}~~,~~\left[\kappa_0 \right] =\frac{T^2}{ML} ~~,~~[d^{\rd}x] = T L^{\rd -1}~.
        \end{split}
\end{align}
We may at times set $c=1$ but expose factors of $c$ when calculating numerical values.
Latin indices take values $a,b,\dots = 0,1,2,\dots, \rd-1$ and Greek indices take values \\ $\mu,\nu,\dots = 0,1,2,\dots, \rd$, with the exception of the Greek letter $\lambda$, which refers to the projective coordinate $x^{\rd} = \lambda = \lambda_0 \ell$.
The covariant derivative acts on contravariant and covariant vectors as
\begin{align}
        \nabla_a V^b = \partial_a V^b + \G^{b}{}_{ac} V^c~~~,~~~\nabla_a V_b = \partial_a V_b - \G^{c}{}_{ab} V_c~~~.
\end{align}
A rank $m$-contravariant, $n$-covariant tensor, which we refer to as an $(m,n)$ tensor, will have $m$-terms involving the connection $\G^a{}_{bc}$ as for contravariant vectors and $n$-terms involving the connection as for covariant vectors. The $\rd+1$-dimensional covariant derivative is defined analogously with $\tG^{\a}{}_{\m\n}$.

At places in this paper where we have assumed compatibility between $\Gamma^{a{}}_{bc}$ and the metric $g_{ab}$ we have
\begin{align}
        \G^{m}{}_{ab} = \frac{1}{2} g^{mn}(g_{n(a,b)} - g_{ab,n})~~~,
\end{align}
but as $G_{\m\n}$ is \emph{never} compatible with $\tilde{\G}^{\a}{}_{\m\n}$, the analogous definition for $\tilde{\G}^{\a}{}_{\m\n}$ is not correct. Instead, $\tilde{\G}^a{}_{mn}$ is defined in Eq.~\eqref{e:Gammatilde}.  The commutator of covariant derivatives on an arbitrary rank $m$-covariant, rank $n$-contravariant tensor is equivalent to the following action of $?R^a_bcd?$
\begin{align}\label{e:RactionGeneral}
        [\nabla_a , \nabla_b ] ?T_{c_1\dots c_m}^{d_1\dots d_n}? =& -?R^e_{c_1 ab}? ?T_{ec_2\dots c_m}^{d_1 d_2 \dots d_n}? - \dots -?R^e_{c_m ab}? ?T_{c_1 c_2\dots e}^{d_1 d_2 \dots d_n}? \cr
        &+ ?R^{d_1}_{eab}? ?T_{c_1\dots c_m}^{e\dots d_n}? + \dots + ?R^{d_m}_{eab}? ?T_{c_1\dots c_m}^{d_1\dots e}? ~~~.
\end{align}
Throughout the paper we adopt the convention that symmetric and antisymmetric permutations of indices do not have numerical factors, i.e.
\begin{align}
    T_{[ab]}&=T_{ab}-T_{ba}\\
    T_{(ab)}&=T_{ab}+T_{ba}~~~.
\end{align}
\section{Reducing the Diffeomorphism Field Transformation Law to One Dimension}\label{a:DtransDetails}
 In this appendix, we demonstrate a detailed proof of the transformation law for the diffeomorphism field $\mathcal{D}$ in one-dimension. As described in section~\ref{s:PathsAndConnections}, under a coordinate transformation from $x^a$ to $x'^a$ in $\rd$-dimensions, the diffeomorphism field $\mathcal{D}_{ab}$ transforms as
\begin{align}
    \mathcal{D}'_{ab} =& \frac{\partial x^m}{\partial x'^a}\frac{\partial x^n}{\partial x'^b} \left[\mathcal{D}_{mn} - \partial_m j_n - j_m j_n + j_c \Pi^c{}_{mn}\right] \;, \\
    j_m \equiv & \partial_m  \log | \tfrac{\partial x^b}{\partial x'^c} |^{\tfrac{1}{\rd+1}} \;.
\end{align}
In $\rd = 1$-dimensions, we have a single field $\mathcal{D}(x)$ and the single connection coefficient $\Pi = 0$ because of its traceless construction as seen in Eq.~\eqref{e:Pi}. The transformation law in $\rd=1$-dimensions is 
\begin{align}\label{e:DTransProof1}
    \mathcal{D}'(x') =& \left(\frac{d x}{d x'}\right)^2 \left[ \mathcal{D}(x) - \frac{d^2}{dx^2} \log \left(\frac{dx}{dx'}\right)^{1/2} - \left(\frac{d}{dx} \log \left(\frac{dx}{dx'}\right)^{1/2}\right)^2\right] \;,
\end{align}
 Considering an infinitesimal coordinate transformation 
\begin{align}
    x' = x - \xi(x) \;,
\end{align}
the diffeomorphism field's transformation law in $\rd=1$-dimensions becomes
\begin{align}\label{e:DTransProof2}
    \mathcal{D}'(x') = \mathcal{D}(x) + 2 \frac{d \xi(x)}{dx} \mathcal{D}(x) - \frac{1}{2} \frac{d^3 \xi(x)}{dx^3} + \mathcal{O}(\xi^2) \;.
\end{align}
The last step is to write $\mathcal{D}'(x')$ in terms of $x$ via a Taylor series
\begin{align}
    \mathcal{D}'(x') =& \sum_{n=0}^{\infty}  \frac{d^n \mathcal{D}'(x')}{dx'} \Bigg|_{x'=x} \frac{(x'-x)^n}{n!} \\
    =& \mathcal{D}'(x) - \xi(x) \frac{d \mathcal{D}'(x)}{dx} + \mathcal{O}(\xi^2) \;.
\end{align}
Plugging this back into the transformation Eq.~\eqref{e:DTransProof2} and rearranging we have
\begin{align}\label{e:DTransProof3}
    \mathcal{D}'(x) = \mathcal{D}(x) + 2 \frac{d \xi(x)}{dx} \mathcal{D}(x) + \xi(x) \frac{d \mathcal{D}'(x)}{dx} - \frac{1}{2} \frac{d^3 \xi(x)}{dx^3} + \mathcal{O}(\xi^2) \;.
\end{align}
Since $\mathcal{D}'(x) = \mathcal{D}(x) + \mathcal{O}(\xi)$, we can simply drop the prime in the third term on the right hand side, leaving us with
\begin{align}\label{e:DTransProof4}
    \mathcal{D}'(x) = \mathcal{D}(x) + 2 \frac{d \xi(x)}{dx} \mathcal{D}(x) + \xi(x) \frac{d \mathcal{D}(x)}{dx} - \frac{1}{2} \frac{d^3 \xi(x)}{dx^3} + \mathcal{O}(\xi^2) \;.
\end{align}
To first order in $\xi$, this is the same as Eq.~\eqref{e:Dtrans1D}.

\section{Conformal Transformations of Curvature Tensors}\label{a:conf}
Here we follow closely~\cite{Dabrowski:2008kx,Kaiser:2010ps} to demonstrate how the Riemann tensor, Ricci tensor, and Ricci scalar transform under a conformal transformation in $\rd$-dimensions:
\begin{align}\label{e:conftrans}
	g_{ab} = e^{-2 \omega} \tilde{g}_{ab}~~~.
\end{align}
This transforms the square root of the determinate as
\begin{align}
	\sqrt{ |g|} = e^{- \rd \omega} \sqrt{|\tilde{g}|}~~~.
\end{align}
The connection $\tilde{\Gamma}^c{}_{ab}$ for the metric $g_{ab}$ and $\Gamma^c{}_{ab}$ for the metric $g_{ab}$ differ by a tensor $\mathcal{C}^c{}_{ab}$ that has the same form for both metrics
\begin{align}
	\Gamma^{c}{}_{ab} =& \tilde{\Gamma}^c{}_{ab} - \mathcal{C}^c{}_{ab} \\
	\mathcal{C}^c{}_{ab} = \tilde{\mathcal{C}}^c{}_{ab} =& \delta_{(a}{}^c \nabla_{b)} \omega - g_{ab} \nabla^c \omega \cr
	=&\delta_{(a}{}^c \tilde{\nabla}_{b)} \omega - \tilde{g}_{ab} \tilde{\nabla}^c \omega~~~.
\end{align}
Here $\nabla_a$ is the covariant derivative associated with the connection $\Gamma^c{}_{ab}$ and  $\tilde{\nabla}_a$ is the covariant derivative associated with the connection $\tilde{\Gamma}^c{}_{ab}$. Indices for the covariant derivatives are raised and lowered with their associated metrics:
\begin{align}
	\nabla^a = g^{ab} \nabla_b~~~,~~~\tilde{\nabla^a} = \tilde{g}^{ab} \tilde{\nabla}_b~~~.
\end{align}
The relationship between the Riemann curvature tensors $R^c{}_{adb}$ constructed from $\Gamma^c{}_{ab}$ and $\tilde{R}^c{}_{adb}$ constructed from $\tilde{\Gamma}^c{}_{ab}$ is
\begin{align}
	R^c{}_{adb} = \tilde{R}^c{}_{adb} - \tilde{\nabla}_{[d} \tilde{\mathcal{C}}^c{}_{b]a} - \tilde{\mathcal{C}}^e{}_{a[d} \tilde{\mathcal{C}}^c{}_{b]e}~~~.
\end{align}
Contracting first and third indices gives the relationship between the Ricci tensors
\begin{align}
	R_{ab} = R^c{}_{acb} = \tilde{R}_{ab} - \tilde{\nabla}_{[c} \tilde{\mathcal{C}}^c{}_{b]a} - \tilde{\mathcal{C}}^e{}_{a[c} \tilde{\mathcal{C}}^c{}_{b]e}
\end{align}
Contracting with the metric and simplifying gives the relationship between the Ricci scalars
\begin{align}
	e^{-2 \omega} R = \tilde{R} + 2 (\rd -1) \tilde{\square} \omega - (\rd -1)(\rd-2) \tilde{\nabla}_a \omega \tilde{\nabla}^a \omega~~~.
\end{align}
Here the Ricci scalars are defined as the Ricci tensor contracted with the associated metric and the Laplacian is defined as contracted with is associated metric:
\begin{align}
	R =& g^{ab} R_{ab}~~~,~~~\tilde{R} = \tilde{g}^{ab} \tilde{R}_{ab} = \tilde{g}^{ab} \tilde{R}^c{}_{acb}~~~\\
	\square \omega =& \nabla^a \nabla_a \omega~~~,~~~\tilde{\square} \omega = \tilde{\nabla}^a \tilde{\nabla_a} \omega~~~.
\end{align}

In proving the above, the following calculations are useful
\begin{align}
	\tilde{g}^{ab} \tilde{C}^c{}_{ab} =& - (\rd -2) \tilde{\nabla}^c \omega \\
	 \tilde{\mathcal{C}}^c{}_{cb} =& \rd \tilde{\nabla}_b \omega \\
	 \tilde{g}^{ad}\tilde{\mathcal{C}}^{c}{}_{ab}\tilde{\mathcal{C}}^b{}_{dc} =& - (\rd -2) \tilde{\nabla}^a \omega \tilde{\nabla}_a \omega~~~.
\end{align}
For a Lagrangian in $\rd$-dimensions of the form $n_\kappa^{-1}\sqrt{|g|} f(\phi) R$ to transform to the Einstein frame
\begin{align}\label{e:EinsteinFrame}
	n_\kappa^{-1}\sqrt{|g|} f(\phi) R =& n_\kappa^{-1}\sqrt{|\tilde{g}|} f(\phi) e^{(2-\rd) \omega}\left( \tilde{R} + 2 (\rd-1) \tilde{\square}\omega - (\rd-1)(\rd-2) \tilde{\nabla}^a \omega \tilde{\nabla}_a \omega \right) \cr
	 =& \sqrt{|\tilde{g}|} \left( \tilde{R} + 2 (\rd-1) \tilde{\square}\omega - (\rd-1)(\rd-2) \tilde{\nabla}^a \omega \tilde{\nabla}_a \omega \right) 
\end{align}
we must have $f(\phi) e^{(2-\rd) \omega} = n_\kappa$, the solution for which is
\begin{align}
	\omega = \tfrac{1}{\rd -2} \ln \tfrac{f(\phi)}{n_\kappa}~~~,~~~\rd \ne 2~~~.
\end{align}
This results in the following for the Laplacian and square of the divergence of $\omega$
\begin{align}
	\tilde{\square} \omega =& \tfrac{1}{\rd-2} \left[\tfrac{f'}{f} \tilde{\square}\phi + \left(\tfrac{f''}{f} - \tfrac{(f')^2}{f^2}\right) \tilde{\nabla}^a \phi \tilde{\nabla}_a \phi \right] \\
	\tilde{\nabla}^a \omega \tilde{\nabla}_a \omega =& \tfrac{1}{(\rd-2)^2} \tfrac{(f')^2}{f^2} \tilde{\nabla}^a \phi \tilde{\nabla}_a \phi
\end{align}
where $f' = df/d\phi$ and $f'' = d^2f/d\phi^2$.
Substituting these results into Eq.~\eqref{e:EinsteinFrame} while multiplying the entire equation by $- M_p^2/2$ gives in terms of $\phi$ now
\begin{align}
	-\tfrac{M_p^2}{2 n_\kappa}\sqrt{|g|} f(\phi) R = -\tfrac{M_p^2}{2}\sqrt{|\tilde{g}|}\left( \tilde{R} + 2 \tfrac{\rd-1}{\rd-2} \tfrac{f'}{f} \tilde{\square}\phi + \tfrac{\rd-1}{\rd-2} \left[2 \tfrac{f''}{f} - 3 \tfrac{(f')^2}{f^2}\right] \tilde{\nabla}^a \phi \tilde{\nabla}_a \phi \right) ~~~.
\end{align}
Under this same conformal transformation, a scalar field Lagrangian transforms as follows
\begin{align}
	\sqrt{|g|} \left[ \tfrac{1}{2} \nabla^a \phi \nabla_a \phi - V(\phi) \right] =& \sqrt{|\tilde{g}|} \left[ \tfrac{1}{2} \tfrac{n_\kappa}{f(\phi)}\tilde{\nabla}^a \phi \tilde{\nabla}_a \phi - \left( \tfrac{n_\kappa}{f(\phi)}\right)^{\tfrac{\rd}{\rd-2}}V(\phi) \right] ~~~.
\end{align}

 \section{Review of Scalar Field Cosmology}\label{a:CosmoReview}     
 \subsection{Friedmann Equations}\label{ss:applications}
The Friedmann-Robertson-Walker (FRW) metric is given by: 
\begin{equation}\label{e:FRW}
    {ds}^2= dt-a(t)^2\left( \frac{dr^2}{1-k{r}^2}+r^2(d\theta+{\sin^2{\theta}}d\varphi) \right)~~~.
\end{equation}
The non-vanishing Christoffel symbols are
\begin{align}
        \Gamma^0{}_{ij} =& - H g_{ij}~~~,~~~\Gamma^i{}_{0j} = H \delta^i{}_j ~~~,~~~\Gamma^1{}_{11} = \tfrac{kr}{1-kr^2} \\
        \Gamma^{\breve{i}}_{1\breve{j}} =& \tfrac{1}{r} \delta^{\breve{i}}{}_{\breve{j}} ~~~,~~~\Gamma^1{}_{\breve{i}\breve{j}} = - \tfrac{1}{r} g^{11}g_{\breve{i}\breve{j}} = \tfrac{1-k r^2}{r a^2}g_{\breve{i}\breve{j}} ~~~,~~~\\
        \Gamma^{2}_{33}  =& - \sin\theta \cos\theta~~~,~~~\Gamma^3{}_{23} = \cot \theta
\end{align}
where the spatial indices are  $i,j,k,\dots  1,2,3$, the spherical indices are $\breve{i},\breve{j},\dots = 2,3$, and the the Hubble parameter is $H(t) = \dot{a}/a$, with an overhead dot indicating a time derivative. The non-vanishing Ricci tensor components and Ricci scalar are
\begin{align}
        R_{00} =& -3 \frac{\ddot{a}}{a} = -3 (\dot{H} + H^2) ~~~,~~~R_{ij} = \frac{1}{3}(R - R_{00}) g_{ij} \\
        R=& -\frac{6}{a^2}(a \ddot{a} + \dot{a}^2 + k) =- 6 \left(\dot{H} + 2 H^2 + \frac{k}{a^2}\right)~~~.
\end{align}
Einstein's equations are~\footnote{Here $\kappa = 8\pi G/c^4$ which is \emph{exactly} equivalent to $M_p^{-2}$ in natural units used throughout the body of the paper. This is not to be confused with $\kappa_0$ which appears throughout the paper and is merely proportional to $M_p^{-2}$ in natural units as shown in Eq.~\ref{e:NattyUnits}.}
\begin{align}\label{e:EE}
        R_{ab} - \frac{1}{2} g_{ab} R - g_{ab}\Lambda_0 = \kappa ~ T_{ab}~~~
\end{align}
where for an FRW metric, the stress-energy tensor plus cosmological constant takes the form of a perfect fluid in co-moving coordinates
\begin{align}
        T_{ab} + g_{ab} \frac{\Lambda_0}{\kappa_0} = \delta_{a}{}^0 \delta_{b}{}^0 (\rho + p) - p g_{ab} ~~~. 
\end{align}
Einstein's equations for an FRW metric, known as the Friedmann equations, are
\begin{subequations}\label{e:FE}
\begin{align}
\label{e:FEa}
        \frac{\ddot{a}}{a} = \dot{H} + H^2 = - \frac{\kappa}{6} (\rho + 3 p) \\
        \label{e:FEE}
        H^2 = \frac{\kappa}{3} \rho - \frac{k}{a^2}~~~.
\end{align}
\end{subequations}
The first equation relates to the acceleration of the scale factor, the second relates the energy density to the critical density $3 H^2/\kappa$.

The energy conservation equation is the derivative of Eq.~\eqref{e:FEE} with Eq.~\eqref{e:FEa} substituted for the acceleration:
\begin{subequations}\label{e:FEECons}
\begin{align}\label{e:Econs}
        \dot{\rho} + 3 H(\rho +p)=0 \\
                \label{e:FEEConsE}
        H^2 = \frac{\kappa}{3} \rho - \frac{k}{a^2}~~~.
\end{align}
\end{subequations}
Simultaneously solving Eqs.~\eqref{e:FEECons} is equivalent to solving the Friedmann equations Eqs.~\eqref{e:FE}. Another form of the Friedmann equations that we will find useful for investigating inflation is found by removing the $H^2$ dependence of the acceleration equation Eq.~\eqref{e:FEa} via the energy equation Eq.~\eqref{e:FEE}:
\begin{subequations}\label{e:FEInf}
\begin{align}\label{e:FEInfa}
	\dot{H} =& - \frac{\kappa}{2} (\rho + p) + \frac{k}{a^2} \\
\label{e:FEInfE}
        H^2 =& \frac{\kappa}{3} \rho - \frac{k}{a^2}~~~.
\end{align}
\end{subequations}
Simultaneously solving Eqs.~\eqref{e:FEInf} is also equivalent to solving the Friedmann equations Eqs.~\eqref{e:FE}.

\subsection{Scalar Field Inflationary Cosmology}
The action for a scalar field is given by\footnote{To avoid confusion, we continue using the notation for the physical inflaton field, $h$, and its potential, $\tilde{V}$.}
\be
S =  \int d^4x\sqrt{-g}\big[\frac{1}{2}\nabla^{m}h\nabla_{m}h-\tilde{V}\big]
\ee
where $\tilde{V}$ is the potential for the scalar field, 
\footnote{in a mostly positive signature metric, the action has an overall negative sign and no relative sign between the kinetic and potential terms.}
noting that in flat space the kinetic term becomes
$\nabla^{m}h\nabla_{m}h= g^{mn}\nabla_{m}h\nabla_{n}h = \dot{h}^2-(\nabla h)^2$.  A massive scalar field has $\tilde{V}=\frac{1}{2}m^2h^2$.
We can find the equations of motion for the scalar field through $\delta S=0$ or equivalently through the Euler-Lagrange Equations,
\be
\frac{\partial \mathcal{L}}{\partial h}-\nabla_{m}\left(\frac{\partial \mathcal{L}}{\partial(\nabla_{m} h)}\right)=0~~~.
\ee
For our scalar field this yields the equation of motion:
\be \label{sfeom}
\nabla^m\nabla_m h+\tilde{V}'=0~~~
\ee
where a prime indicates differentiation with respect to $h$
\begin{align}
    \tilde{V}' = \frac{d \tilde{V}}{dh}~~~.
\end{align}
A massive scalar field's Klein-Gordon equation is $\nabla^m\nabla_m h+m^2 h=0$.
The Stress-energy tensor is in general defined by
\be
T^{mn} = -\frac{2}{\sqrt{-g}}\frac{\delta S}{\delta g_{mn}} = -\frac{2}{\sqrt{-g}}\frac{\delta(\sqrt{-g}\mathcal{L})}{\delta g_{mn}} = -g^{mn}\mathcal{L} -2\frac{\delta \mathcal{L}}{\delta g_{mn}}
\ee
so the scalar field stress energy tensor is given by
\be
T_{\phi}^{mn}=-g^{mn}\big[\frac{1}{2}\nabla^{a} h \nabla_{a} h-\tilde{V}\big]+\nabla^m h\nabla^n h~~~.
\ee
The stress-energy tensor for a perfect fluid in general coordinates is given by
\be
T^{mn} = (\rho+p)u^{m}u^{n}-p g^{mn}~~~.
\ee
This matches to a perfect fluid if we set
\begin{align}
\rho_{h} =& \frac{1}{2}\nabla^{a}h\nabla_{a} h+\tilde{V} \\
p_{h}=& \frac{1}{2}\nabla^{a} h\nabla_{a} h-\tilde{V} \\
u^{m}=&\frac{\nabla^{m} h}{\sqrt{ \nabla^{a} h\nabla_{a} h}}
\end{align}
where $g_{mn}u^{m}u^{n}=1$ is satisfied for our conventions.

If we treat the field to be homogeneous, i.e. $h(t)$, then the Lagrangian reduces to
\be
\mathcal{L}=\frac{1}{2}\dot{h}^2-\tilde{V}
\ee
and the energy density and pressure become
\begin{subequations}\label{e:prho}
\begin{align}
 \label{e:rhophi}
\rho_{h} =& \frac{1}{2}\dot{h}^2+\tilde{V}\\
\label{e:pphi} 
p_{h}=& \frac{1}{2}\dot{h}^2-\tilde{V}~~~.
\end{align}
\end{subequations}
In the Friedman-Robertson-Walker metric, Eq.~\eqref{e:FRW}, 
the equation of motion Eq.~\eqref{sfeom} becomes
\be\label{e:phiEQM}
\tilde{V}'=-3\frac{\dot{a}}{a}\dot{h}-\ddot{h} = -3H\dot{h}-\ddot{h} 
\ee
which is proportional to the energy conservation equation Eq.~\eqref{e:Econs}. Thus solving the Friedmann equations automatically solves the scalar field equation as well.  We can see an additional term compared to the equation of motion of a scalar field in a Minkowski background, which can be thought of as a friction term due to the expansion of the universe.

The energy density $\rho$ in the Friedmann Equations represents the energy density contribution of all relevant forms of energy.  In general, this would be $\rho=\rho_{r}+\rho_{m}+\rho_{\Lambda} + \rho_{\text{other}}$.  In the case of the scalar field acting as an inflaton field, it would be the dominant energy density of the universe during inflation, so $\rho=\rho_{h}$.
Plugging $\rho_{h}$ and $p_{h}$ from Eqs.~\eqref{e:prho} into Eqs.~\eqref{e:FEInf} gives
\begin{subequations}\label{e:FEInfrhop}
\begin{align}\label{e:FEInfarhop}
	\dot{H} =& - \frac{\kappa}{2} \dot{h}^2 + \frac{k}{a^2} \\
\label{e:FEInfErhop}
        H^2 =& \frac{\kappa}{3}(\tfrac{1}{2} \dot{h}^2 + \tilde{V}) - \frac{k}{a^2}~~~.
\end{align}
\end{subequations}
The slow roll parameter is defined as the negative of the ratio of these two equations
\be
\epsilon_H\equiv -\frac{\dot{H}}{H^2}~~~.
\ee
The defining condition of inflation is that it was a period of accelerated expansion of the universe, corresponding to $\ddot{a}>0$.  From the acceleration equation Eq.~\eqref{e:FEa}, $\ddot{a}>0$ implies that $H^2 > - \dot{H}$ and thus $\epsilon_H<1$.  

As the universe expands, $a$ increases and therefore the $\frac{k}{a^2}$ term becomes negligible compared to the energy density of the scalar field.  Then we have that
\begin{subequations}
\begin{align}
\dot{H}=&-\frac{\kappa}{2} \dot{h}^2
 \\
 \label{e:FEInfEk0}
H^2=&\frac{\kappa(\frac{1}{2}\dot{h}^2+\tilde{V})}{3}  
\end{align}
\end{subequations}
so the slow roll parameter becomes
\be
\epsilon_H = \frac{3}{1+2\tfrac{\tilde{V}}{\dot{h}^2}}~~~.
\ee
Then the condition $\epsilon_H<1$ implies $\tilde{V} > \dot{h}^2$ so inflation requires a potential dominated expansion.  If $\tilde{V} >> \dot{h}^2$, this has the consequences that $p\simeq -\rho$ coming from Eqs.~\eqref{e:prho} and the Freidmann Equation Eq.~\eqref{e:FEInfEk0} and the slow roll parameter $\epsilon_H$ becomes
\begin{align}\label{e:FriedmannEnergyApprox}
	H^2 \simeq & \kappa \tilde{V}/3 \\
	\label{e:epsilonApprox}
	\epsilon_H \simeq & \frac{3 \dot{h}^2}{2 \tilde{V}} ~~~.
\end{align}

In most inflationary scenarios it is assumed that the field acceleration is negligible, i.e. $\ddot{h}\sim 0$, corresponding to the field `slowly rolling.'  Then the field equation of motion Eq.~\eqref{e:phiEQM} simplifies to
\be\label{e:ScalarEQMSlowRoll}
3H\dot{h}\simeq-\tilde{V}'~~~.
\ee
Substituting $H$ from Eq.~\eqref{e:FriedmannEnergyApprox} into Eq.~\eqref{e:ScalarEQMSlowRoll} results in the following equation for the scalar field
\be\label{phiV}
\dot{h}=-\frac{\tilde{V}'}{\sqrt{3\kappa \tilde{V}}}~~~.
\ee
Substituting into Eq.~\eqref{e:epsilonApprox}, the slow roll parameter can be written purely in terms of the potential,
\be\label{eV}
\epsilon_H \simeq \frac{1}{2\kappa}\bigg(\frac{\tilde{V}'}{\tilde{V}}\bigg)^2<1~~~.
\ee
Differentiating Eq.~\eqref{eV} gives a further constraint
\be
\eta\equiv \frac{1}{\kappa}\bigg(\frac{\tilde{V}''}{\tilde{V}}\bigg)<1~~~. 
\ee
These two conditions on the potential amount to requiring it is flat in the sense that it has small derivatives.  A chosen model of inflation then amounts to choosing a $\tilde{V}$ such that these conditions are satisfied.  Some example potentials are
\begin{itemize}
\item   $\tilde{V}=\frac{1}{2}m^2h^2$\quad Massive scalar field
\item  $\tilde{V}=\lambda h^4$\quad Self-interacting scalar field (A special case of polynomial inflation $\tilde{V}\propto h^{\alpha}$)
\item $\tilde{V}=\lambda(h^2-M^2)^2$\quad Higgs potential 
\item $\tilde{V}\propto e^{c h}$\quad Power law inflation
\item $\tilde{V}=\frac{1}{4\lambda}(M^2-\lambda\psi^2)^2+U+\frac{1}{2}g^2\psi^2 h^2$\quad Hybrid inflation with two fields
\end{itemize}
A further constraint on models of inflation is the duration. The number of \emph{e}-folds $N$ is defined as
\begin{align}
	N \equiv \ln\left( \frac{a}{a_0} \right) = \int_{t_*}^{t_{end}} H dt~~~.
\end{align} 
The last equality can be derived using the definition of $H = \dot{a}/a$. Performing a change of variables $dt = dh/\dot{h}$ and utilizing the approximate Eq.~\eqref{e:FriedmannEnergyApprox} and Eq.~\eqref{e:ScalarEQMSlowRoll}, the number of \emph{e}-folds can be cast into the following approximate form,
\begin{align}
	N \simeq \frac{1}{M_p^2} \int_{h_\text{end}}^{h_*} \frac{\tilde{V}}{\tilde{V'}} dh~~~.
\end{align}
There must be approximately 60 $e$-foldings (though values of 50-70 can be found without the inflation literature) and the model must incorporate an end to inflation in accordance with observation.
For further information see \cite{Liddle:1999mq,Weinberg:2008zzc}.

\bibliographystyle{JHEP}
\bibliography{Bibliography}
\end{document}